\documentclass[a4paper,conference,10pt]{IEEEtran}
\usepackage[ruled,vlined]{algorithm2e}
\usepackage[T1]{fontenc} 
\usepackage[utf8]{inputenc}

\usepackage{amsfonts,epsfig,cite,array,multirow,graphicx,amsmath,ltablex,tabularx,setspace,amssymb,multirow}
\usepackage{schemabloc,amsthm}
\usepackage{subfig}
\usepackage[numbers,sort&compress]{natbib}
\usetikzlibrary{circuits, arrows}
\usepackage{verbatim}
\usepackage{graphicx}
\usepackage{mathtools}
\usepackage{xcolor}
\usepackage{marvosym}

\usepackage{lipsum}
\usepackage{dblfloatfix}
\usepackage{color}
\usepackage{etoolbox}
 \usepackage{epsfig}
\usepackage[inline]{enumitem}
\theoremstyle{plain}

\newtheorem{rem}{Remark}
\newtheorem{exm}{Example}

\makeatletter

\begin{document}
\SetKwRepeat{Do}{do}{while}
\newcolumntype{L}[1]{>{\raggedright\let\newline\\\arraybackslash\hspace{0pt}}m{#1}}
\newcolumntype{C}[1]{>{\centering\let\newline\\\arraybackslash\hspace{0pt}}m{#1}}
\newcolumntype{R}[1]{>{\raggedleft\let\newline\\\arraybackslash\hspace{0pt}}m{#1}}
\newcommand{\oeq}{\mathrel{\text{\sqbox{$=$}}}}
\title{
\huge{Design of  SCMA Codebooks  Using Differential Evolution}}

\author{Kuntal Deka$^{1}$, Minerva Priyadarsini$^{1}$, Sanjeev Sharma$^{2}$, and Baltasar Beferull-Lozano (Senior Member, IEEE)$^{3}$\\
	$^1$Indian Institute of Technology Goa, India,   $^2$Indian Institute of Technology (BHU) Varanasi, India\\
	$^3$Department of Information
and Communication Technology, University of Agder, Grimstad 4879,
Norway} 

\maketitle

\begin{abstract}
Non-orthogonal  multiple access (NOMA) is a promising technology which meets the demands of  massive connectivity in future wireless networks. Sparse code multiple access (SCMA) is  a popular  code-domain NOMA technique.  The effectiveness of SCMA comes from:  (1) the multi-dimensional sparse codebooks  offering high shaping gain and (2) sophisticated  multi-user detection based on message passing algorithm (MPA).   The codebooks of the users play the main role in determining the performance of SCMA system. 
This paper presents a framework to design the  codebooks by taking into account the entire system including the SCMA encoder and the MPA-based detector. The  symbol-error rate (SER) is considered as the  design criterion which  needs to be minimized.  Differential evolution (DE) is used to carry out the minimization of the SER over the codebooks. The codebooks of the SCMA  systems of overloading factor 150$\%$ and 200$\%$ are designed and displayed in the paper. The simulation results are presented for various channel models.
\end{abstract}

\textbf{Keywords}
SCMA, codebook design, differential evolution, message passing algorithm.


\section{Introduction}

The future wireless communication system will comprise  of  enormous number of interconnected  devices.
The   massive connectivity required   in such situations cannot be  fulfilled by the  multiple access schemes deployed so far. Earlier,  3G system used code division multiple access (CDMA), 4G used orthogonal frequency division multiple access (OFDMA) \cite{dai2015non}. All these technologies are based on orthogonal multiple access (OMA) principle. In OMA, the number of supported users is limited by the number of available orthogonal resources. The non-orthogonal  multiple access (NOMA) technology  provides a gateway for massive connectivity.  It supports overloaded systems where  the number of users is higher than the number of orthogonal resources.  NOMA techniques can be broadly classified into two groups: \textit{power domain} and \textit{code domain}.   In power domain NOMA,  different levels of powers are  allocated to different users.  
In code domain NOMA, different codewords or signatures  are used for different users. Low-density spreading (LDS) is a code-domain NOMA technique where,  a user's symbol
is  multiplied with a distinct  sparse spreading
signature.  This spread-ed sequence is mapped to modulation constellation points  for transmission \cite{hoshyar_lds}.   Nikoupor \textit{et al.}  came up with the  technique of  sparse code multiple access (SCMA)  in order to improve upon LDS \cite{nikopour2013}. In SCMA, the operations of the spreading and the modulation mapping are merged. Based on a dedicated codebook, a symbol is directly  mapped to a sparse multi-dimensional codeword. Thus, SCMA provides a better opportunity  than LDS to  attain  high shaping gain.
Owing to the inherent sparsity in the code-domain NOMA, the multi-user detection is usually done  by the message passing algorithm (MPA) \cite{spa}.
{\underline{\textit{Related work}}}:
The performance of an SCMA system is mainly determined by the codebooks of the users.  The optimization of the codebooks for SCMA system is a convoluted  task as multiple users are interfering with multi-dimensional complex vectors.  In   \cite{nikopour2013,taherzdeh2014}, first, the optimum codebook for one user is designed. The remaining codebooks were obtained  by carrying out    user-specific operations on the optimized codebook.
In   \cite{zhang2016capacity} (referred to  as ``Zhang \cite{zhang2016capacity}"),   the sum-rate was considered as one of the design criteria. First, a   series of one-dimensional complex codewords were designed.
Then the angles of these codewords were changed with the objective of  improving the sum-rate.
The authors in \cite{alam2017designing} considered the   minimum Euclidean distance and the energy diversity of the constellation points  to obtain the optimum  codebooks.
Yu \textit{et al.} (referred to  as ``Yu \cite{star}") designed    SCMA codebooks  based on  the star quadrature amplitude modulation (QAM) constellations \cite{star}.
In \cite{Bao_2018TCOM}, the authors designed  codebooks for  bit-interleaved convolutionally-coded SCMA system.
This design was   aided by the analysis based on EXtrinsic Information Transfer (EXIT) chart. Sharma \textit{et al.} \cite{sharma_globecom} (referred to as ``Sharma \cite{sharma_globecom}")  designed the codebooks by maximizing the mutual information and shaping gain. Kim \textit{et al.} proposed a deep-learning-based method where, the codebook is dynamically designed with the objective of minimizing the bit error rate (BER) \cite{kim_deep}.

{\underline{\textit{Contributions}}}:
 The existing codebook-design  methods  mainly focus on various geometric properties of the multi-dimensional constellations with little emphasis on  detection.  It is difficult to track the MPA-based detection process mathematically.  The factor graph is finite,  not tree-like and it contains short cycles. Due to these reasons, the techniques like density evolution and EXIT chart cannot accurately characterize the MPA-based multi-user detection process.
We propose to consider the symbol error rate (SER)  as the cost function which is to be minimized.  The SER is one such quantity which  takes into account every part of the multi-user system be it Euclidean distance profile, product distance profile, MPA etc. The reliability of the system is precisely reflected by the SER.  We adopt differential evolution (DE) for the minimization of the SER as it is not a simple function of the codebooks.  DE is a flexible and effective evolutionary algorithm which is used  to solve complex optimization problems with real-valued parameters \cite{Storn1997, DE_2011}.  First, the structure of the codebooks is represented  with the help of a finite number of constellation points. Then a DE-based optimization process is invoked to find  the optimum constellation points. 
The  codebooks for the additive white Gaussian noise (AWGN)  and Rayleigh fading channels are designed.   The SER performance of the proposed codebooks are compared with those of the existing ones.  Moreover, various key parameters of the codebooks are computed and analyzed.

 {\underline{\textit{Outline}}}:   Section~\ref{sec::prelim} describes the  preliminaries such as SCMA system model, important parameters of codebooks and DE. The proposed method of codebook design is presented in Section~\ref{sec::prop}. The simulation results are presented and analyzed in Section~\ref{sec::simulations}.  Section~\ref{sec::conc} concludes the paper.
\section{Preliminaries} \label{sec::prelim}
\subsection{System Model} \label{sec::sys_model}
A $J\times K$ SCMA system refers to an overloaded  multi-user scenario with $J$ users and   $K$ resource elements. The overloading factor is given by $\lambda=\frac{J}{K}$.
A dedicated codebook ${\cal{C}}_j$  containing  $M$ $K$-dimensional codewords: ${\cal{C}}_j= \left\{{\bf{x}}_{j1}, {\bf{x}}_{j2}, \ldots, {\bf{x}}_{jM}\right\}$ is assigned to every $j^{\text{th}}$ user.
Each codeword ${\mathbf{x}}_{jm}$  is  sparse with  $N$ non-zero complex components.
Based on the assigned codebook ${\cal{C}}_j$,  $\log_{2}(M)$  data bits are directly mapped to the $K$-dimensional codeword $\mathbf{x}_{j}=\left[x_{j1},...,x_{jK}\right]^T$.
The received signal ${\mathbf{y}}=\left[y_1, \ldots y_K\right]^T$ is:
\begin{equation}\label{int1}
\mathbf{y}=\sum_{j=1}^{J} {\text{diag}}\left({\bf{h}}_j\right)\mathbf{x}_{j}+\mathbf{n}
\end{equation}
where, ${\bf{h}}_j=\left[h_1, \ldots, h_K\right]^T$ is the  channel gain vector for the $j^{\text{th}}$ user   and $\bf{n}$ is a complex  $K \times 1$ AWGN vector.
\begin{figure}[htb!]
	\centering
	\scalebox{1}{\includegraphics[]{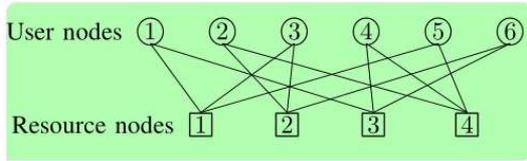}}
	\caption {SCMA block diagram. }	 	
	\label{fig::factor_graph1}
\end{figure}
The locations of the non-zero elements of the codewords for the users can be represented with the help of a factor matrix as shown in (\ref{F_mat}).  The  $1$s present in the $j$th column  specify the  locations of the non-zero components of the codewords for the $j$th user.
The matrix $F$ can be alternatively represented by a factor graph as shown in Fig.~\ref{fig::factor_graph1}. The degree
of a resource node is denoted by $d_f$. This specifies that $d_f$ users interfere with each others over one resource element. As the factor graph is sparse,  MPA is  used for multi-user detection.
\begin{equation}
\label{F_mat}
{\bf{F}}=
\begin{bmatrix}
1 & 0 & 1 & 0 & 1 & 0\\
0 & 1 & 1 & 0 & 0 & 1\\
1 & 0 & 0 & 1 & 0 & 1\\
0 & 1 & 0 & 1 & 1 & 0\\
\end{bmatrix}
\end{equation}

 The performance of an SCMA system is highly sensitive to the codebooks. In literature \cite{boutros98,codebook_design_survey},  various key performance  indicators (KPIs)  have been  considered for the design of multi-dimensional constellations.  A few of them are highlighted:\\
	$\bullet$ {\textit{{Minimum Euclidean Distance ($d_{E,\min}$)}}}:  It is defined as the minimum distance  between any pair $\left({\bf{x}}_m, {\bf{x}}_n\right)$ of  codewords in the entire SCMA system:
		$$ d_{E,\min}=\min_{m,n} ||{\bf{x}}_m-{\bf{x}}_n||.$$
	$\bullet$ {\textit{{Euclidean kissing number ($\tau_{E}$)}}}:   It is defined as the number of distinct codeword pairs with Euclidean distance equal to $d_{E,\min}$.\\
	$\bullet$ {\textit{{Minimum product distance ($d_{P,\min}$)}}}:  The  product distance $d_P^{m,n}$ between two codewords ${\bf{x}}_m$ and ${\bf{x}}_n$ is defined as
	 $$d_P^{m,n}=\prod_{j\in {\cal{J}}_{m,n}}|x_{mj}-x_{nj}|$$
	 	where,  ${\cal{J}}_{m,n}$ is  the set of dimensions for which $x_{mj}\neq x_{nj}$. The minimum product distance $d_{P,\min}$ is given by: $$ d_{P,\min}=\min_{m,n} d_P^{m,n}.$$
	$\bullet$ {\textit{{Product Kissing Number ($\tau_{P}$)}}}:  It is defined as the number of distinct codeword pairs with product distance equal to $d_{P,\min}$.

The objective is to maximize $d_{E,\min}$, $d_{P,\min}$  and minimize $\tau_{E}$, $\tau_{P}$. In addition to  these parameters, mutual information between the received signal and the sum of the interfering codewords is also a vital parameter.  Suppose $Y$ is the signal received over one resource element. Let $S$ represent  the sum of the codewords of the $d_f$  users interfering over that resource.  We have a total of  $M^{d_f}$   distinct sum values  for $S$  as given by
$\left\{s_1, s_2, \ldots, s_{M^{d_f}}\right\}$. A high value of $I(Y;S)$ ensures successful recovery of the individual user's data from the noisy  sum value.
Usually a lower bound $I_L$ on $I(Y;S)$ is considered\footnote{$I(Y;S)$ or $I_L$ is related to the sum-rate \cite{zhang2016capacity}.}.  $I_L$ for AWGN channel with noise variance $N_0$ is given by \cite{zhang2016capacity}
\begin{equation}
\label{lower_bound}
\resizebox{0.47 \textwidth}{!}{$I_L= \log_2 M^{d_f} -
 \log_2 \left[1+\frac{1}{M^{d_f}} \sum_{j=1}^{M^{d_f}}
\sum_{\substack{i=1 \\ i \neq j}}^{M^{d_f}}\exp \left(-\frac{1}{4N_0} |s_j-s_i|^2\right)\right]$}.
\end{equation}

\subsection{Differential Evolution}
In DE, an initial set of $S_P$ random candidate solutions or vectors  $\{{\mathbf{p}}_i: i = 1,2,\ldots, S_P\}$ is generated. The length of each vector is the same and it  is denoted by $D$.  New generations are created iteratively  using mutation, cross over and selection process.  During mutation, each vector ${}\mathbf{p}_i^G$, where $G$ denotes the generation index, is selected as a primary parent. For each parent, a mutant vector is generated as follows:
\begin{equation}
\mathbf{u}_i^G={\mathbf{p}}_{r_1}+\alpha(\mathbf{p}_{r_2}-\mathbf{p}_{r_3})
\end{equation}
where,  $r_1, r_2, r_3 $ are randomly selected distinct numbers different from $i$ and $\alpha$ is the scaling factor.  $\mathbf{u}_i^G$ is the secondary parent generated from mutation. Cross-over is applied to primary and secondary parent to obtain the offspring vector $\mathbf{v}_i^G$. Each component $v_{ik}^G$ of $\mathbf{v}_i^G = [ v_{i1}^G,\ldots,v_{ik}^G, \ldots,  v_{iD}^G]$ is inherited from either the primary or the secondary parent as per the following rule:
\begin{align*}
v_{ik}^G=
\begin{cases}
u_{ik}^G,  \text { if $h_k \leq C_r$}\\
p_{ik}^G,  \text { otherwise}
\end{cases}
\end{align*}
where, $h_k$ is a random number  uniformly distributed over [0,1], $C_r$ is the cross-over rate.
The offspring  is made to inherit at least one component from the secondary parent to ensure that  offspring is different from primary parent.
 The offspring vector $\mathbf{v}_i^G$ has to compete with the parent vector $\mathbf{p}_k^G$ to get a place in the next $G+1$ generation. If $\mathbf{v}_k^G$ gives a lower cost function than $\mathbf{p}_k^G$ then the former replaces the later, else the primary parent vector exists in the next generation too, i.e.
\begin{align*}
\mathbf{p}_i^{(G+1)}=
\begin{cases}
\mathbf{v}_i^G , \text { if $f(v_i^G)\leq f(p_i^G)$}\\
\mathbf{p}_i^G, \text{ otherwise}
\end{cases}
\end{align*}
where $f(\cdot)$ is the objective function that needs to be minimized. The new generation performs at least as good as the best candidate of the previous generation. These steps are continued  until some stopping criteria are  satisfied. The best vector from the current population is considered as the optimum vector.
\section{Proposed Codebook Design based on Differential Evolution}  \label{sec::prop}
Let ${\bf{C}}=\left\{{\cal{C}}_1, \ldots, {\cal{C}}_J\right\}$ denote the set or the collection of the codebooks of all the users.
The task of SCMA codebook design is formulated as an optimization problem as follows:
\begin{equation}
\begin{aligned}
{\bf{C}}_{\text{opt}}=&\arg  \min_{{\bf{C}}}  f\left({\bf{C}}; \frac{E_b}{N_0}, {\bf{F}}\right)\\
&{\text{s.t.}} \;\; ||{\bf{c}}||_2=1 \;\;\forall \bf{c} \in {{\bf{C}}}
\end{aligned}
\label{opt}
\end{equation}
where, $f\left(\cdot\right)$ is the SER at SNR $\frac{E_b}{N_0}$ and   $\bf{F}$  is the factor graph matrix.  Every $K$-dimensional codeword ${\bf{c}}$ present in the SCMA codebook system $\bf{C}$ is constrained to have unit Euclidean  norm. 



Note that an already-designed  factor graph matrix $\bf{F}$    is fed to the optimization process in (\ref{opt}).
Suppose the  factor graph is  regular with the  resource node degree of $d_f$. Then, $d_f$ users overlap on a particular resource node. One needs to carefully assign  constellation points to these users. The task is to  assign  $d_f$  distinct one-dimensional constellation\footnote{One-dimensional codebook is assigned as  only one resource node is considered at a time.} codebooks to the $d_f$ overlapping users.  These codebooks may be  represented by $C_1^o, C_2^o, \ldots, C_{d_f}^o$, where the superscript `$o$' signifies that these are one-dimensional.  The size of the constellation is  $M$. For example,   consider the case of $6\times$4 SCMA system with the factor graph shown  in Fig.~\ref{fig::factor_graph1} and the matrix in (\ref{F_mat}). Here,  $d_f=3$ and $M=4$. The three  one-dimensional codebooks can be represented as follows:

\begin{equation}
\begin{gathered}
C_1^o=[a_1\;\;   a_2\;  -a_2 \; -a_1] \\
C_2^o=[a_3 \;  \;a_4\;  -a_4 \; -a_3] \\
C_3^o=[a_5\;  \; a_6\;  -a_6\;  -a_5]
\end{gathered}
\label{one_dim}
\end{equation}
where,  $a_i \in \mathbb{C}$, $i=1,\ldots, 6$.

In (\ref{one_dim}),  12 distinct complex numbers are created from 6 complex numbers and their negative counterparts. It is preferable to design the SCMA codebooks with the  minimum possible  number of constellation points.  This  reduces the hardware requirement in implementation.

The one-dimensional codebooks can be assigned to all the resource nodes through a structure matrix satisfying Latin property
 \cite{latin}.   This way of generating  the entire set of the codebooks   was considered in \cite{zhang2016capacity,xiao_capacity}.  The structure matrix for the $6\times 4$ SCMA system is shown below:
 \begin{equation}
 \label{F_mat_str}
 {\bf{F}}_L=
 \begin{bmatrix}
 C_1^o & 0 & C_2^o & 0 & C_3^o & 0\\
 0 & C_2^o & C_3^o & 0 & 0 & C_1^o\\
  C_2^o & 0 & 0 & C_1^o & 0 & C_3^o\\
 0 & C_1^o & 0 & C_3^o & C_2^o & 0\\
 \end{bmatrix}
 \end{equation}

 The Latin property in (\ref{F_mat_str})  ensures that distinct one-dimensional codebooks are assigned to the users overlapping over any particular resource.
 Using ${\bf{F}}_L$, the set of all codebooks can be generated as follows:
 \begin{equation}
 \begin{gathered}
 {\cal{C}}_1=\begin{bmatrix}
 C_1^o\\
 {\bf{0}} \\
 C_2^o \\
 {\bf{0}}
 \end{bmatrix} \;\;
 {\cal{C}}_2=\begin{bmatrix}
 {\bf{0}} \\
 C_2^o \\
 {\bf{0}}\\
 C_1^o\\
 \end{bmatrix} \;
 {\cal{C}}_3=\begin{bmatrix}
 C_{2,p}^o\\
 C_3^o \\
 {\bf{0}} \\
 {\bf{0}}
 \end{bmatrix} \\
 {\cal{C}}_4=\begin{bmatrix}
 {\bf{0}}\\
 {\bf{0}} \\
 C_1^o\\
 C_3^o \\
 \end{bmatrix} \;
 {\cal{C}}_5=\begin{bmatrix}
 C_3^o\\
 {\bf{0}}\\
 {\bf{0}} \\
 C_{2,p}^o \\

 \end{bmatrix} \;
 {\cal{C}}_6=\begin{bmatrix}
 {\bf{0}} \\
 C_1^o\\
 C_3^o \\
 {\bf{0}}
 \end{bmatrix}
 \end{gathered}
 \label{compact_codebook}
 \end{equation} 
 where, $\bf{0}$ is the all-zero row vector of length $M=4$.

 In order to obtain more diversity and higher shaping gain, the authors in \cite{zhang2016capacity,xiao_capacity} have proposed to carry out the dimensional permutation switching algorithm (DPSA). As per DPSA,  the permuted version of $C^o_2$ denoted by $C^o_{2,p}$ is used in the codebooks ${\cal{C}}_3$  and ${\cal{C}}_5$ in place of $C^o_2$. We consider $C^o_{2,p}=\left[ -a_4,\; a_3, -a_3,\; a_4\right]$.
%
With this dimensional permutation, using (\ref{one_dim})   in (\ref{compact_codebook}), the codebooks can be written as shown in TABLE~\ref{table:codebook_structure}.
 \begin{table}[ht]
 	\caption{Structure  $\mathbb{S}$ of the codebooks ${\bf{C}}=\left\{{\cal{C}}_1, \ldots, {\cal{C}}_6\right\}$}
 	\label{table:codebook_structure}
 \scalebox{0.8}{
 	$
 \begin{gathered}
 {\cal{C}}_1=\left\{\begin{bmatrix}
 a_1 \\
 0\\
 a_3\\
 0
 \end{bmatrix}
 \begin{bmatrix}
 a_2 \\
 0\\
 a_4\\
 0
 \end{bmatrix}
 \begin{bmatrix}
 -a_2 \\
 0\\
 -a_4\\
 0
 \end{bmatrix}
 \begin{bmatrix}
 -a_1 \\
 0\\
 -a_3\\
 0
 \end{bmatrix}  \right\}  \;
 {\cal{C}}_2=\left\{\begin{bmatrix}
 0\\
 a_3\\
 0\\
 a_1
 \end{bmatrix}
 \begin{bmatrix}
 0\\
 a_4\\
 0\\
 a_2
 \end{bmatrix}
 \begin{bmatrix}
 0\\
 -a_4\\
 0\\
 -a_2
 \end{bmatrix}
 \begin{bmatrix}
 0\\
 -a_3\\
 0\\
 -a_1
 \end{bmatrix}   \right\} \\
 {\cal{C}}_3=\left\{\begin{bmatrix}
 -a_4\\
 a_5\\
 0\\
 0
 \end{bmatrix}
 \begin{bmatrix}
  a_3\\
 a_6\\
 0\\
 0
 \end{bmatrix}
 \begin{bmatrix}
 -a_3\\
- a_6\\
 0\\
 0
 \end{bmatrix}
 \begin{bmatrix}
 a_4\\
 -a_5\\
 0\\
 0
 \end{bmatrix}   \right\}  \;
 {\cal{C}}_4=\left\{\begin{bmatrix}
 0\\
 0\\
 a_1\\
 a_5
 \end{bmatrix}
 \begin{bmatrix}
 0\\
 0\\
 a_2\\
 a_6
 \end{bmatrix}
 \begin{bmatrix}
0\\
0\\
-a_2\\
-a_6
 \end{bmatrix}
 \begin{bmatrix}
0\\
0\\
-a_1\\
-a_5
 \end{bmatrix}   \right\}\\
  {\cal{C}}_5=\left\{\begin{bmatrix}
 a_5\\
 0\\
 0\\
 -a_4
 \end{bmatrix}
 \begin{bmatrix}
 a_6\\
 0\\
 0\\
 a_3
 \end{bmatrix}
 \begin{bmatrix}
 -a_6\\
 0\\
 0\\
 -a_3
 \end{bmatrix}
 \begin{bmatrix}
 -a_5\\
 0\\
 0\\
 a_4
 \end{bmatrix}   \right\} \;
 {\cal{C}}_6=\left\{\begin{bmatrix}
 0\\
 a_1\\
 a_5\\
 0
 \end{bmatrix}
 \begin{bmatrix}
0\\
a_2\\
a_6\\
0
 \end{bmatrix}
 \begin{bmatrix}
0\\
-a_2\\
-a_6\\
0
 \end{bmatrix}
 \begin{bmatrix}
 0\\
 -a_1\\
 -a_5\\
 0
 \end{bmatrix}   \right\}
 \end{gathered}$}
\end{table}
 Observe that the codebooks for the $6\times 4$ SCMA system can be represented with the help of 6 complex numbers $a_i$, $i=1,\ldots, 6$. These complex numbers can be compactly represented by ${\boldsymbol{a}}=\left[a_1, \ldots, a_6\right]$. Any particular complex vector ${\boldsymbol{a}}_i$, ${\boldsymbol{a}}_i \in {\mathbb{C}}^6$ refers to a particular set of codebooks ${\mathbf{C}}_i=\left\{{\cal{C}}^i_1, \ldots, {\cal{C}}^i_6\right\}$. With these notations, the SCMA codebook design problem in (\ref{opt}) can now be written as
  \begin{equation}
  {\boldsymbol{a}}_{\text{opt}}=\arg  \min_{{\boldsymbol{a}}}  f\left({\boldsymbol{a}}; \frac{E_b}{N_0}\right)
  \label{opt1}
  \end{equation}
 where, $f\left({\boldsymbol{a}}; \frac{E_b}{N_0}\right)$ is the SER of the SCMA system defined by $\boldsymbol{a}$ at SNR $\frac{E_b}{N_0}$. In the rest of the paper, for notational brevity, $f\left({\boldsymbol{a}}; \frac{E_b}{N_0}\right)$ will be represented as $f\left({\boldsymbol{a}}\right)$ with the understanding that the SNR is fixed at a particular value during the optimization process.  In (\ref{opt1}), ${\boldsymbol{a}}_{\text{opt}}$ is that vector $\boldsymbol{a}$ which yields the optimum codebook ${\bf{C}}_{\text{opt}}$ as per the structure given in TABLE~\ref{table:codebook_structure}.  Moreover, we have the constraint that the Euclidean norm of every codeword is 1 although it is not explicitly mentioned in (\ref{opt1}).

\begin{algorithm}[!htbp]
	\caption{\small Codebook design using differential evolution}
	\label{algo:codebook_DE}
	\small
	\SetKwData{Left}{left}
	\SetKwData{This}{this}
	\SetKwData{Up}{up}
	\SetKwFunction{Union}{Union}
	\SetKwFunction{FindCompress}{FindCompress}
	\SetKwInOut{Input}{input}
	\SetKwInOut{Output}{output}
	\Input{Codebook structure $\mathbb{S}$ (TABLE~\ref{table:codebook_structure}),  $\frac{E_b}{N_0}$,  DE parameters ($\alpha$,  $C_r$, $S_P$ and $I_{\max}$)}
	\Output{Codebooks ${\bf{C}}_{\text{opt}}=\left\{{\cal{C}}_1, \ldots, {\cal{C}}_6\right\}$. }
	\BlankLine
	%
	Initialize the population matrix ${\bf{P}}$ to a  matrix of size $S_P \times D$ having random numbers uniformly distributed over $[-1,1]$ where $D=12$, Normalize these entries so that the every codeword of the codebook has unit norm\;
	
	\While{termination criteria not fulfilled}{
		\For{$i\leftarrow 1$ \KwTo $S_P$}{	
			\textcolor{black}{Select three distinct vectors (rows) ${\bf{p}}_{r_0}$,${\bf{p}}_{r_1}$ \text{ and } ${\bf{p}}_{r_2}$ uniformly at random from $\bf{P}$ such  that they are also different from ${\bf{p}}_i$}\;
			Generate an integer $j_{\text{rand}}$ uniformly at  random from $\left\{1,2, \ldots, D \right\}$\;
			\tcc{\scriptsize{Generation of trial vector ${\bf{u}}$}}
			\For{$j\leftarrow 1$ \KwTo $D$}{
				\If{${\text{rand[0,1]}} \leq C_r$  or $j=j_{\text{rand}}$ }{
					$\textcolor{black}{{{u}}_{j,i}}={{p}}_{j,r_0}+\alpha\times\left({{p}}_{j,r_1}-{{p}}_{j,r_2}\right)$ \tcp*[f]{\scriptsize Crossover and Mutation}
					
				}
				\Else{
					${{u}}_{j,i}={{p}}_{j,i}$
				}	
				
			}	
		Normalize the entries of $\bf{u}$ so that every codeword has unit norm.
		
			\tcc{\scriptsize{Evaluation and Selection}}	
			
		    Suppose ${\bf{\mathbf{C}}}^{\bf{u}}$ and   ${\bf{\mathbf{C}}}^{{\bf{p}}_i}$ are the two  sets  of the codebooks for all users as per $\bf{u}$  and ${\bf{p}}_i$ respectively.
		
			Run Monte Carlo simulation for the SCMA system with the codebooks ${\bf{\mathbf{C}}}^{\bf{u}}$ and   ${\bf{\mathbf{C}}}^{{\bf{p}}_i}$ at SNR $\frac{E_b}{N_0}$. Suppose,        $f({\bf{u}})$ and $f({\bf{p}}_i)$ are the respective   SER values     \;
			\If{$f({\bf{u}}) <f({\bf{p}}_i)$}{
				${\bf{p}}_i={\bf{u}}$  \tcp*[f]{\scriptsize{Replace the $i$th row of $\bf{P}$ by $\bf{u}$}}\;
			}		
		}
		From the updated population matrix $\bf{P}$, find the vector (row) ${\bf{p}}_{\text{opt}}$   which yields the minimum value  of SER\;
		If $f\left({\bf{p}}_{{\text{opt}}}\right)$ is not changing significantly from the previous iteration or the maximum number of iterations $I_{\max}$ are exhausted, then break  from  loop\;
	}
	
	Using ${\bf{p}}_{\min}$,  form the optimum constellation vector ${\boldsymbol{a}}_{\text{opt}}=\left[a_1,\ldots, a_6\right]$.   Then inserting  ${\boldsymbol{a}}$ in $\mathbb{S}$,  the optimized set of the codebooks  of all users ${\mathbf{C}}_{\text{opt}}=\left\{{\cal{C}}_1, \ldots, {\cal{C}}_J\right\}$ are obtained.  \;	
\end{algorithm}

We solve (\ref{opt1}) with the help DE. The job is to find the optimum vector ${\boldsymbol{a}}$ which contains 6 complex numbers\footnote{It is noteworthy that the proposed method of codebook design is universal in the sense that it can be applied to an SCMA system with any values of $J$ and $K$.  The structure as shown in TABLE~\ref{table:codebook_structure} must be fed to the DE-based optimizer. }.  However,  DE is a real-valued optimization technique. Therefore, in this case, the number of real-valued variables to be optimized is $D=12$.  The detailed steps for the DE-based codebook design are presented in Algorithm~\ref{algo:codebook_DE}. The inputs to the algorithm  are  the structure $\mathbb{S}$ for the codebooks as shown in TABLE~\ref{table:codebook_structure}, SNR $\frac{E_b}{N_0}$ and the DE parameters: $\alpha$,  $C_r$,$S_P$ and $I_{\max}$ .
The candidate vectors are stored in a population matrix $\bf{P}$ in   row-wise manner. The total number of candidate vectors is $S_P$. 
The size of $\bf{P}$  is $S_P\times D$.   The elements of  $\bf{P}$ are initialized to random numbers uniformly distributed over  $\left[-1,1\right]$.  Every row ${\bf{p}}_s$, $s=1, \ldots, S_P$,  of $\bf{P}$ corresponds to  an SCMA codebook system specified by the 6 complex numbers ${\boldsymbol{a}}_s=\left[a_{s,1}, \ldots, a_{s,6}\right]$. Suppose $a_{s,t}=a_{s,t}^r+a_{s,t}^ci$, $t=1,\ldots, 6$, where $a_{s,t}^r$ and $a_{s,t}^c$ are real uniform numbers in $[-1,1]$. The $s^{\text{th}}$ row   ${\bf{p}}_s$ of ${\bf{P}}$ is given by
$$
{\bf{p}}_s=\left[a_{s,1}^r, a_{s,1}^c, a_{s,2}^r, a_{s,2}^c, \ldots ,a_{s,6}^r, a_{s,6}^c \right].
$$
The elements of $\bf{P}$  are normalized so that  every codeword in  a codebook system has unit norm. Against  each row ${{\bf{p}}_s}, s = 1, \ldots , S_P$, a trial vector $\bf{u}$ is generated with the given values of the DE parameters: crossover rate ($C_r$) and
scaling factor ($\alpha$). Algorithm~\ref{algo:codebook_DE} shows the  detailed steps of the generation of the trial vector $\bf{u}$.   The decision regarding the replacement of the current population vector ${\bf{p}}_s$ by the  trial vector $\bf{u}$  is made by computing the SER values for the two SCMA systems through Monte Carlo simulations. If the SER value $f\left(\bf{u}\right)$ of the SCMA system defined by $\bf{u}$ is less than the SER $f\left({\bf{p}}_i\right)$ of the system  defined     by  ${\bf{p}}_i$, then the  current population  vector ${\bf{p}}_i$ is set to $\bf{u}$. In this way, every candidate vector in $\bf{P}$ is examined
and updated with the trial vectors if necessary. From the updated population $\bf{P}$, the best vector ${\bf{p}}_{\text{opt}}$ with the minimum SER value is identified. The above-mentioned process is repeated unless $f\left({\bf{p}}_{\text{opt}}\right)$  is not changing significantly during two consecutive iterations  or the maximum number of iterations are exhausted. The real numbers in  ${\bf{p}}_{\text{opt}}$ converted to the corresponding complex numbers to obtain ${\boldsymbol{a}}_{\text{opt}}$.
The complex numbers in ${\boldsymbol{a}}_{\text{opt}}=\left[a_1,\ldots, a_6\right]$ are plugged into   $\mathbb{S}$ in TABLE~\ref{table:codebook_structure} to obtain the optimum codebooks.

\begin{exm}
	Consider the case of $J=6$, $K=4$ SCMA system with constellation size $M=4$. The structure given in TABLE~\ref{table:codebook_structure} is used.  The number of variables is $D=12$. We can consider $S_P=20$, $C_r=0.95$ and $\alpha=0.6$.  The population matrix $\bf{P}$ may be initialized to the  following  $20\times 12$ matrix:
\begin{center}	
\scalebox{0.6}
{$
{\bf{P}}=\left[ {\begin{array}{rrrrrrrrrrrr}
	0.46 &  -0.53 & -0.90 &  -0.22 &  -0.68 &  0.16 &  -0.34 & -0.08 &  0.22 &  0.91 &  0.49 &  0.48\\
	-0.02 & 0.90 &   0.44 &  0.08 & -0.44 & 0.01 & 0.32 & -0.84 & 0.10 & 0.24 &   0.82 &  0.12 \\
	-0.65 & -0.62 &  0.80 &  0.11 & - 0.41  & -0.15  & -0.55 & 0.22 & 0.56 & 0.22&    0.41  & 0.15\\
	\cdot & \cdot &\cdot &\cdot &\cdot &\cdot &\cdot &\cdot &\cdot &\cdot &\cdot &\cdot \\
	\cdot & \cdot &\cdot &\cdot &\cdot &\cdot &\cdot &\cdot &\cdot &\cdot &\cdot &\cdot \\
	\cdot & \cdot &\cdot &\cdot &\cdot &\cdot &\cdot &\cdot &\cdot &\cdot &\cdot &\cdot \\
	\end{array} }\right]. $}
\end{center}
Due to space constraint, only 3 rows out of $20$ are shown.   It may be verified  that the norm of  every codeword in $\bf{P}$   is 1.  For every row of $\bf{P}$, a trial vector is generated by carrying out the mutation and the crossover operations as specified in Algorithm~\ref{algo:codebook_DE}. If the trial vector yields less SER than the current row vector, then the current row is overwritten by the trial vector. In this way every row of $\bf{P}$ is examined and updated if needed. Suppose,  finally,  the population matrix ${\bf{P}}$ is as given below  where the first row is the best row ${\bf{p}}_{\min}$.
\begin{center}	
	\scalebox{0.7}
	{$
		{\bf{P}}=\left[ {\begin{array}{rrrrrrrrrrrr}
			-0.33 &  0.63 & -0.83  & 0.43 & 0.71 & 0 &  -0.36 & 0 & -0.42&  - 0.84 &    0.59 & 0.35\\			
			\cdot & \cdot &\cdot &\cdot &\cdot &\cdot &\cdot &\cdot &\cdot &\cdot &\cdot &\cdot \\
		\cdot & \cdot &\cdot &\cdot &\cdot &\cdot &\cdot &\cdot &\cdot &\cdot &\cdot &\cdot \\
			\cdot & \cdot &\cdot &\cdot &\cdot &\cdot &\cdot &\cdot &\cdot &\cdot &\cdot &\cdot \\
			\end{array} }\right]. $}
\end{center}
Then the   solution of (\ref{opt1}) is given by the following:
\vspace{-0.2in}
\begin{center}	
	\scalebox{0.75}
	{
$
{\boldsymbol{a}}_{\rm{opt}}=\left[-0.33+0.63i,  -0.83+0.43i,  0.71,  -0.36,  -0.42- 0.84i,     0.59+0.35i \right].
$}
\end{center}
Using the above ${\boldsymbol{a}}_{\rm{opt}}$ in TABLE~\ref{table:codebook_structure}, we can generate the codebooks for the SCMA system. TABLE~\ref{table:awgn} shows  these codebooks   in Section~\ref{sec::simulations}.
\end{exm}

\begin{rem}
	Although the proposed scheme
	appears to be SNR-dependent, the codebooks optimized at a  high SNR are
	found to work well at different SNR values. In practice, we fix an SNR where the SER is around $10^{-3}$. The results for different SNR values are not shown here due to space constraint. 
\end{rem}


\textit{\underline{Complexity Analysis}}:
Observe from Algorithm~\ref{algo:codebook_DE} that during each cycle of DE, the objective functions (SERs) for the current population and the trial vector are computed.  The SER computation is a time-consuming process due to fact that the complexity of the MPA is $O\left(M^{d_f}\right)$.  Thus, the complexity of proposed algorithm is  higher than some of the existing codebook designing methods.  However, note that the codebook design is a one-time and    offline task. It does not add to the real-time complexity of the system. Therefore, the proposed DE-based codebook design method is feasible in practical scenario.
%
\section{Simulation Results} \label{sec::simulations}

The simulations are carried out  to evaluate the performance of the proposed codebooks. For comparison,  the following  codebooks are considered:
\begin{itemize*}
	\item ``Zhang \cite{zhang2016capacity}"
	\item  ``Sharma \cite{sharma_globecom}"
	\item ``Yu \cite{star}"
	\item  ``Ma \cite{NOMA_book}" (Chapter 12 of \cite{NOMA_book}).
\end{itemize*}
As per the thumb rules provided in \cite{DE_price_20,qing_DE}, the DE parameters are set to: $S_P=20, D=12,  C_r=0.95, \alpha=0.6$ and $I_{\max}=80$.  The simulations are done for   uncoded  transmission over AWGN and
flat Rayleigh fading  channels.
\subsection{AWGN channel}
\begin{figure}[htb!]
	\centering
	\includegraphics[width=2.55in ,height=2.5in]{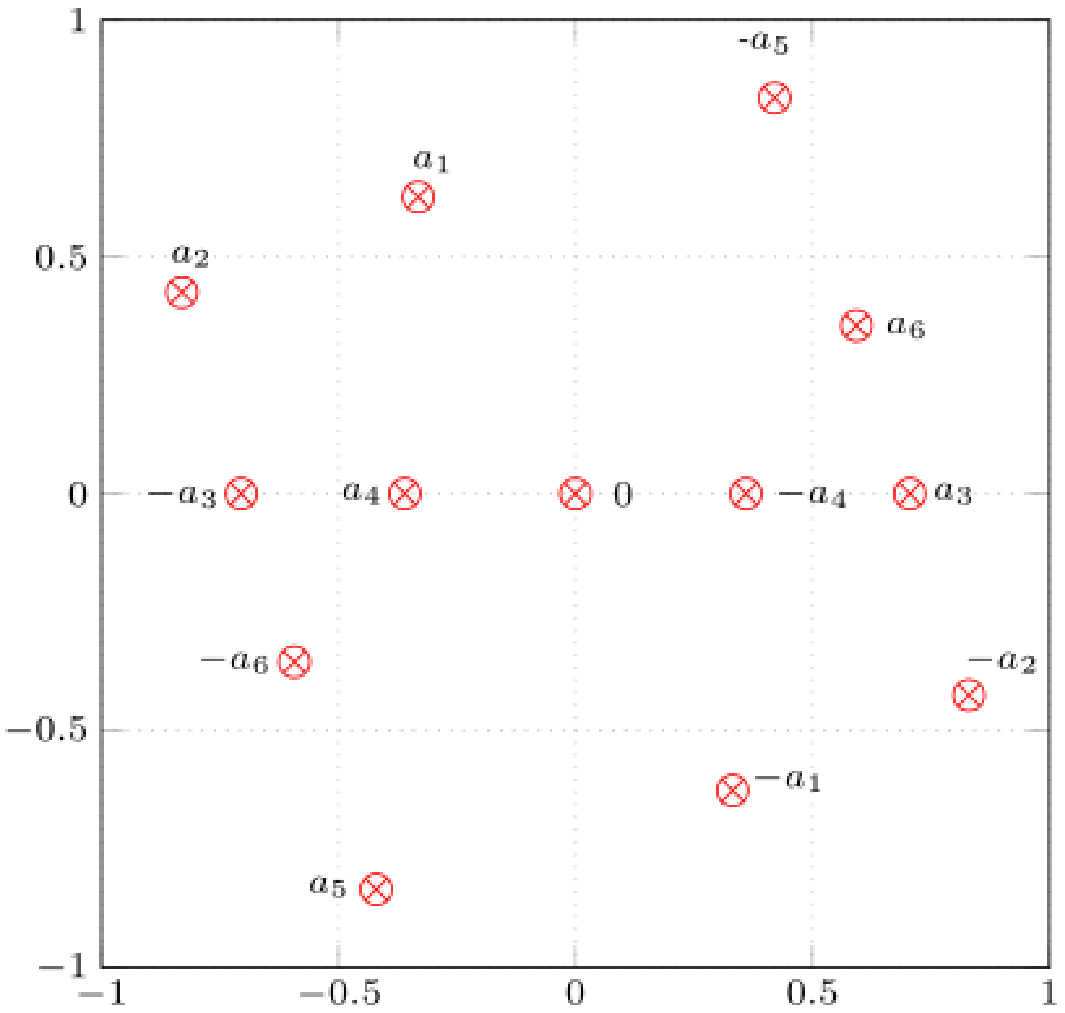}
	\caption {${\boldsymbol{a}}_{\text{opt}}$ for AWGN channel. }	 	
	\label{const_awgn}
\end{figure}

 First the results for the AWGN channel are presented.  An SCMA system with   $J=6$, $K=4$  and $M=4$ is considered.   
 Algorithm~\ref{algo:codebook_DE} is run at $\frac{E_b}{N_0}=10$ dB to find the optimum vector ${\boldsymbol{a}}_{\text{opt}}$. The complex numbers of ${\boldsymbol{a}}_{\text{opt}}$ and their negative versions are shown in Fig.~\ref{const_awgn}.
  With this value of ${\boldsymbol{a}}_{\text{opt}}=\left[a_1,a_2,a_3,a_4,a_5,a_6\right]$, the optimum codebooks are generated using the structure $\mathbb{S}$ shown in TABLE~\ref{table:codebook_structure}.  These codebooks are shown in TABLE~\ref{table:awgn}.
  \begin{table}[ht]
 	\caption{Codebooks optimized for AWGN channel}
 	\label{table:awgn}
 	
\scalebox{0.6}{
 $
 \begin{gathered}
  {\cal{C}}_1=\left\{\begin{bmatrix}
  -0.3318 + 0.6262i\\
  0 \\
  0.7055 \\
  0
  \end{bmatrix}
  \begin{bmatrix}
  -0.8304 + 0.4252i \\
  0  \\
  -0.3601  \\
  0
  \end{bmatrix}
   \begin{bmatrix}
  0.8304 - 0.4252i \\
  0  \\
  0.3601  \\
  0
  \end{bmatrix}
  \begin{bmatrix}
  0.3318 - 0.6262i\\
  0 \\
  -0.7055 \\
  0
  \end{bmatrix}  \right\} \\
  {\cal{C}}_2=\left\{\begin{bmatrix}
  0\\
  0.7055\\
  0\\
  -0.3318 + 0.6262i
  \end{bmatrix}
  \begin{bmatrix}
   0\\
  -0.3601 \\
  0\\
  -0.8304 + 0.4252i
  \end{bmatrix}
  \begin{bmatrix}
  0\\
  0.3601 \\
  0\\
  0.8304 - 0.4252i
  \end{bmatrix}
 \begin{bmatrix}
 0\\
 -0.7055\\
 0\\
 0.3318 - 0.6262i
 \end{bmatrix}   \right\} \\
 {\cal{C}}_3=\left\{\begin{bmatrix}
 0.3601 \\
 -0.4202 - 0.8350i\\
 0\\
 0
 \end{bmatrix}
 \begin{bmatrix}
 0.7055\\
 0.5933 + 0.3548i \\
 0\\
 0
 \end{bmatrix}
 \begin{bmatrix}
 -0.7055\\
 -0.5933 - 0.3548i \\
 0\\
 0
 \end{bmatrix}
 \begin{bmatrix}
 -0.3601 \\
 0.4202  +0.8350i\\
 0\\
 0
 \end{bmatrix}   \right\} \\
 {\cal{C}}_4=\left\{\begin{bmatrix}
 0\\
 0\\
-0.3318 + 0.6262i\\
-0.4202 - 0.8350i
 \end{bmatrix}
 \begin{bmatrix}
 0\\
 0\\
 -0.8304 + 0.4252i\\
 0.5933 + 0.3548i
 \end{bmatrix}
 \begin{bmatrix}
  0\\
 0\\
 0.8304 - 0.4252i\\
 -0.5933 - 0.3548i
 \end{bmatrix}
 \begin{bmatrix}
 0\\
 0\\
 0.3318 - 0.6262i\\
 0.4202 + 0.8350i
 \end{bmatrix}   \right\} \\
 {\cal{C}}_5=\left\{\begin{bmatrix}
 -0.4202 - 0.8350i \\
 0\\
 0\\
 0.3601
 \end{bmatrix}
 \begin{bmatrix}
  0.5933 + 0.3548i\\
0\\
0\\
0.7055
 \end{bmatrix}
 \begin{bmatrix}
 -0.5933 - 0.3548i\\
 0\\
 0\\
 -0.7055
 \end{bmatrix}
 \begin{bmatrix}
 0.4202 +0.8350i \\
 0\\
 0\\
 -0.3601
 \end{bmatrix}   \right\} \\
  {\cal{C}}_6=\left\{\begin{bmatrix}
  0\\
-0.3318 + 0.6262i\\
-0.4202 - 0.8350i\\
0
 \end{bmatrix}
 \begin{bmatrix}
 0\\
 -0.8304 + 0.4252i\\
 0.5933 + 0.3548i\\
 0
 \end{bmatrix}
 \begin{bmatrix}
  0\\
 0.8304 - 0.4252i\\
 -0.5933 - 0.3548i\\
 0
 \end{bmatrix}
 \begin{bmatrix}
 0\\
0.3318 - 0.6262i\\
0.4202 + 0.8350i\\
0
 \end{bmatrix}   \right\}
 \end{gathered}
$}
\end{table}
Observe that Euclidean norm of every codeword is 1.

\begin{figure}[htb!]
	\centering
	\scalebox{1}{\includegraphics[]{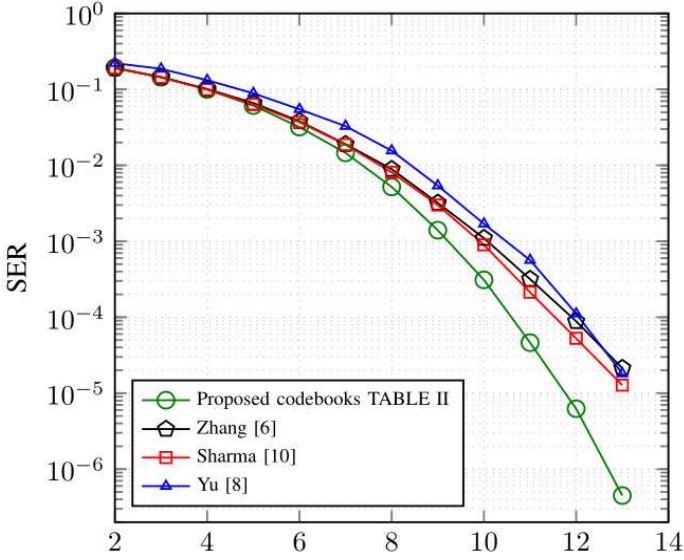}}
	\caption {SER performance of the SCMA system using various codebooks for  $J=6$ and $K=4$  in AWGN channel. }	 	
	\label{6by4}
\end{figure}
The SER performance of the proposed codebooks are evaluated through Monte Carlo simulations  and these are shown in Fig.~\ref{6by4}.  The SER plots for the other codebooks are also presented.  Observe that the proposed codebooks outperform the others by a significant margin. Specifically, there is coding gain of about 1.4 dB at SER=$10^{-5}$  over the next best codebooks (``Sharma~\cite{sharma_globecom}"). The performance of ``Ma \cite{NOMA_book}" over AWGN channel is not satisfactory. Therefore, its SER plot is not shown in Fig.~\ref{6by4}. However,  ``Ma \cite{NOMA_book}"  yields  good results over Rayleigh fading channel   which is presented later in this section.
\begin{figure}[htb!]
	\centering
	\scalebox{0.6}{\includegraphics[]{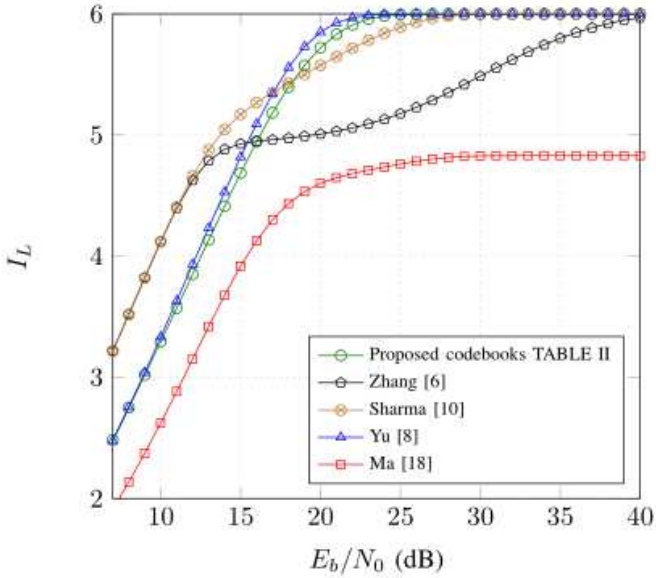}}
	\caption {Lower bound on the mutual information for various codebooks.}	 	
	\label{cap}
	\vspace{-0.15in}
\end{figure}
The SER performance of the various codebooks can be studied and justified by carrying out the  mutual information analysis as mentioned in Section~\ref{sec::sys_model}. We plot the lower bound $I_L$ on mutual information for various codebooks in Fig.~\ref{cap}.  Observe that in the SNR region below 16 dB, ``Shamra \cite{sharma_globecom}" and ``Zhang \cite{zhang2016capacity}" provide higher values of  $I_L$ than the proposed codebooks.  However, the $I_L$ for  ``Yu \cite{star}" and  the proposed codebooks reach the maximum value of 6 quicker than the other codebooks.  Observe that $I_L$ for  ``Yu \cite{star}" is slightly higher than that for the proposed codebooks in the range of 15-20 dB. They reach the maximum value almost at the same SNR of 24 dB.  However, as described later in this section, the Euclidean distance and the product distance  profiles of ``Yu \cite{star}"  are poorer than those of the proposed one.  This observation reinforces the superior performance of the proposed DE-based codebooks. Also note  that the $I_L$ for ``Ma \cite{NOMA_book}"  cannot climb up to the maximum value.   This justifies the poor performance of
``Ma \cite{NOMA_book}" over the AWGN channel.

\subsection{Fading channel}
\begin{figure}[htb!]
	\centering
	\includegraphics[width=2.55in ,height=2.5in]{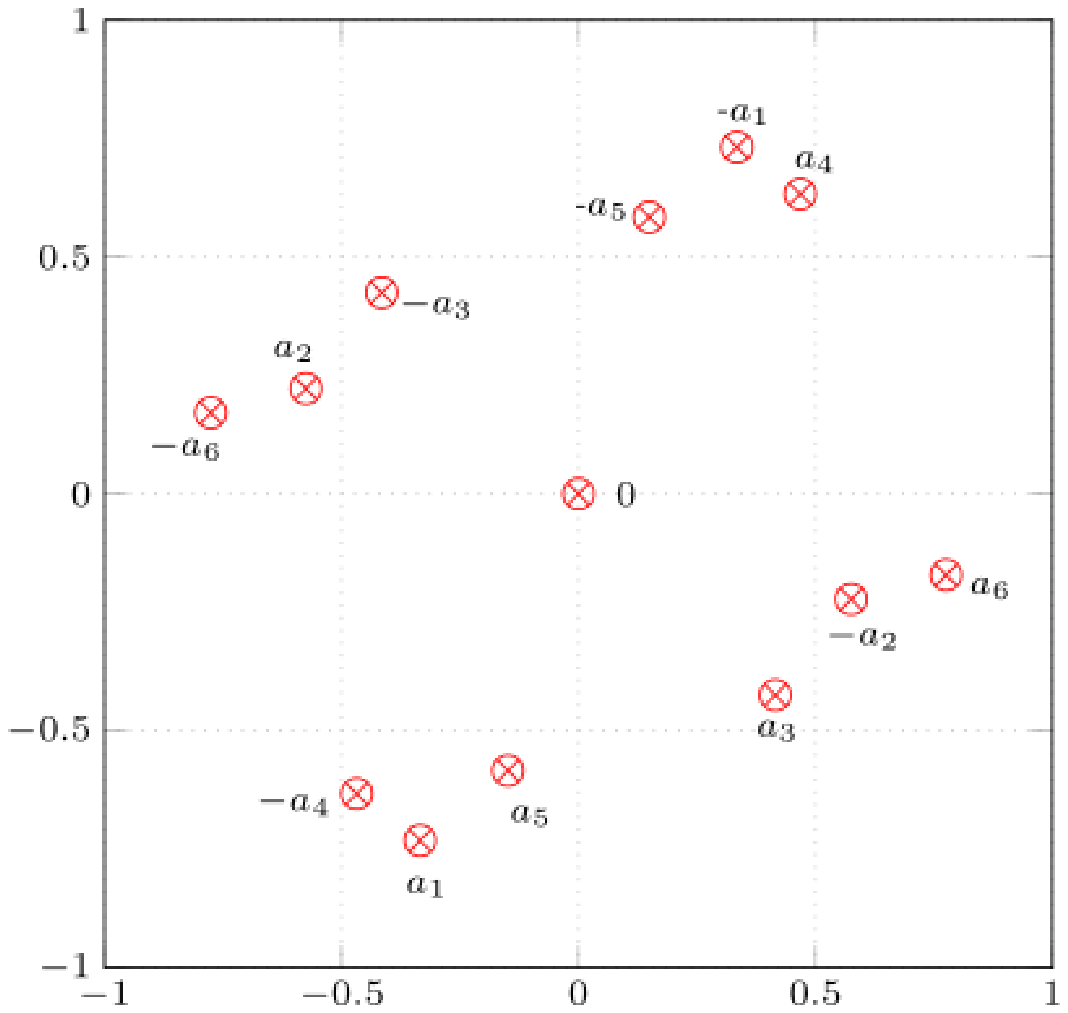}
	\caption {${\boldsymbol{a}}_{\text{opt}}$ for  fading channel.}	 	
	\label{const_fading}
\end{figure}

In this case, each user observes independent Rayleigh fading channel coefficients over the resource elements. The same SCMA framework with  $J=6$, $K=4$,  $M=4$  and the  structure $\mathbb{S}$
given in TABLE~\ref{table:codebook_structure} is considered.  
Algorithm~\ref{algo:codebook_DE} is executed at $\frac{E_b}{N_0}=17$ dB. It yields the optimum constellation vector ${\boldsymbol{a}}_{\text{opt}}$   which are plotted in Fig.~\ref{const_fading}.
The complex numbers in ${\boldsymbol{a}}_{\text{opt}}=\left[a_1, \ldots, a_6\right]$ are put in $\mathbb{S}$.  The  resulting codebooks are shown in TABLE~\ref{table:fading}.  Observe that the optimized codebooks for the fading channel are different from those for the AWGN channel. This signifies that the codebook design problem depends on the underlying channel model.
\begin{table}[ht]
	\caption{Codebooks optimized for fading channel}
	\label{table:fading}
	
	\scalebox{0.6}{
		$
		\begin{gathered}
		{\cal{C}}_1=\left\{\begin{bmatrix}
		-0.3344 - 0.7316i \\
		0\\
		0.4153 - 0.4248i\\
			0
		\end{bmatrix}
		\begin{bmatrix}
		-0.5754 + 0.2224i \\
		0\\
		0.4680 + 0.6328i\\
			0
		\end{bmatrix}
		\begin{bmatrix}
		0.5754 - 0.2224i \\
			0\\
		-0.4680 - 0.6328i\\
		0
		\end{bmatrix}
		\begin{bmatrix}
		0.3344 + 0.7316i \\
		0\\
		-0.4153 + 0.4248i\\
		0
		\end{bmatrix}  \right\} \\
		{\cal{C}}_2=\left\{\begin{bmatrix}
		0\\
		0.4153 - 0.4248i\\
		0\\
		-0.3344 - 0.7316i
		\end{bmatrix}
		\begin{bmatrix}
			0\\
		0.4680 + 0.6328i\\
		0\\
		-0.5754 + 0.2224i
		\end{bmatrix}
		\begin{bmatrix}
		0\\
		-0.4680 - 0.6328i\\
		0\\
		0.5754 - 0.2224i
		\end{bmatrix}
		\begin{bmatrix}
		0\\
		-0.4153 + 0.4248i\\
		0\\
		0.3344 + 0.7316i
		\end{bmatrix}   \right\} \\
		{\cal{C}}_3=\left\{\begin{bmatrix}
		-0.4680 - 0.6328i\\
		-0.1492 - 0.5839i\\
		0\\
		0
		\end{bmatrix}
		\begin{bmatrix}
		0.4153 - 0.4248i\\
		0.7759 - 0.1713i \\
			0\\
			0
		\end{bmatrix}
		\begin{bmatrix}
		-0.4153 + 0.4248i\\
		-0.7759 + 0.1713i\\
			0\\
			0
		\end{bmatrix}
		\begin{bmatrix}
		0.4680 + 0.6328i\\
		0.1492 + 0.5839i\\
			0\\
			0
		\end{bmatrix}   \right\} \\
		{\cal{C}}_4=\left\{\begin{bmatrix}
			0\\
			0\\
		-0.3344 - 0.7316i\\
		-0.1492 - 0.5839i
		\end{bmatrix}
		\begin{bmatrix}
			0\\
			0\\
		-0.5754 + 0.2224i \\
		0.7759 - 0.1713i
		\end{bmatrix}
		\begin{bmatrix}
			0\\
			0\\
		0.5754 - 0.2224i \\
		-0.7759 + 0.1713i
		\end{bmatrix}
		\begin{bmatrix}
			0\\
			0\\
		0.3344 + 0.7316i\\
		0.1492 + 0.5839i
		\end{bmatrix}   \right\} \\
		{\cal{C}}_5=\left\{\begin{bmatrix}
		-0.1492 - 0.5839i \\
			0\\
			0\\
		-0.4680 - 0.6328i
		\end{bmatrix}
		\begin{bmatrix}
		0.7759 - 0.1713i\\
			0\\
			0\\
		0.4153 - 0.4248i
		\end{bmatrix}
		\begin{bmatrix}
		-0.7759 + 0.1713i\\
			0\\
			0\\
		-0.4153 + 0.4248i
		\end{bmatrix}
		\begin{bmatrix}
		0.1492+0.5839i \\
			0\\
			0\\
		0.4680 + 0.6328i
		\end{bmatrix}   \right\} \\
		{\cal{C}}_6=\left\{\begin{bmatrix}
			0\\
		-0.3344 - 0.7316i \\
		-0.1492 - 0.5839i \\
			0
		\end{bmatrix}
		\begin{bmatrix}
			0\\
		-0.5754 + 0.2224i \\
		0.7759 - 0.1713i \\
			0
		\end{bmatrix}
		\begin{bmatrix}
			0\\
		0.5754 - 0.2224i \\
		-0.7759 + 0.1713i \\
			0
		\end{bmatrix}
		\begin{bmatrix}
			0\\
		0.3344 + 0.7316i \\
		0.1492 + 0.5839i \\
			0
		\end{bmatrix}   \right\}
		\end{gathered}
		$}
\end{table}


\begin{figure}[htb!]
	\centering
	\scalebox{1}{\includegraphics[]{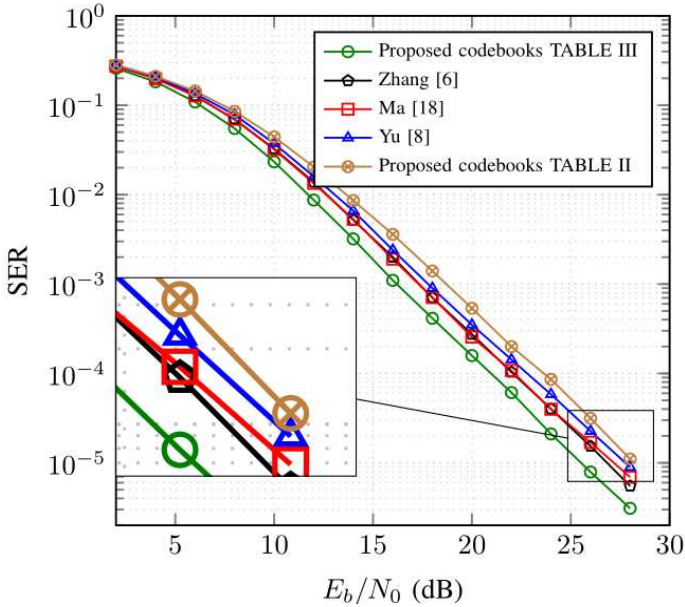}}
	\caption {SER performance of the SCMA system using various codebooks for  $J=6$ and $K=4$  in Rayleigh fading channel.}	 	
	\label{6by4_fading}
\end{figure}
The SER performance of the proposed codebooks along with those of the other existing ones are shown in Fig.~\ref{6by4_fading}.  Observe that the proposed codebooks produce the best results. At SER=$10^{-5}$, we experience a  coding gain of  about 1.5 dB over the next best methods: ``Zhang~\cite{zhang2016capacity}" and ``Ma~\cite{NOMA_book}". We also evaluate the SER performance of the proposed codebooks designed for the AWGN channel. However, the performance is not satisfactory and inferior to  ``Zhang~\cite{zhang2016capacity}", ``Ma~\cite{NOMA_book}" and ``Yu \cite{star}". This observation reiterates the well known fact that the optimum constellation for AWGN  may not be optimum for Rayleigh fading channel and vice versa \cite{boutros98}.

\vspace{-0.1in}
\begin{figure}[htb!]
	\centering
	\includegraphics[width=2.55in ,height=2in]{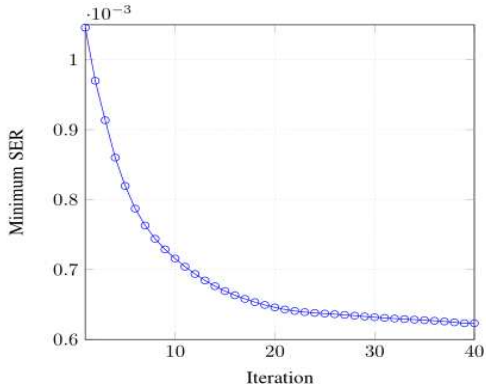}
	\caption {Minimum SER versus iteration  in Rayleigh fading channel at $\frac{E_b}{N_0}=17$ dB during differential evolution process.}	 	
	\label{{min_ser_fading}}
\end{figure}

The progress of Algorithm~\ref{algo:codebook_DE} with increasing  iteration  is depicted in Fig.~\ref{{min_ser_fading}}. The minimum SER corresponding to the best
vector/row in the  population matrix $\bf{P}$ as updated in the current iteration  is plotted against the iteration number.  Observe from Fig.~\ref{{min_ser_fading}} that the minimum SER tends to settle down at a value  after 20  iterations.  We have observed similar progression of the DE-based algorithm in the case of AWGN channel. However, due to space constraint, the plot for AWGN channel is not included in the paper.

The values of the KPIs mentioned in Section~\ref{sec::sys_model}  are shown in TABLE~\ref{table::codebook_pararmeters} for various codebooks.  ``Zhang~\cite{zhang2016capacity}" has the best Euclidean distance  profile. Its $d_{E,\min}$ is the highest and $\tau_{E}$  is the lowest.   Proposed codebook (AWGN) has the highest $d_{P,\min}$, however the $\tau_P$ is not the lowest. The Euclidean distance parameters for ``Ma~\cite{NOMA_book}" are the worst since $d_{E,\min}$ is the lowest and $\tau_{E}$  is the highest. Its poor performance over AWGN channel can be attributed to this fact. However, its product distance profile is impressive.  Its $d_{P,\min}$ is not the lowest and $\tau_{P}$ has the lowest value. However, it is difficult to justify  the reported SER performances completely with the help of the KPIs mentioned in TABLE~\ref{table::codebook_pararmeters}. 

\begin{table}[h]
	\centering
	\caption{ Key performance indicators}
	\label{table::codebook_pararmeters}
	\scalebox{0.78}{
		\begin{tabular}{|c|C{1.3cm}|C{1.3cm}|c|c|c|c|}
			\hline
			 & \small{Proposed (AWGN)} &  \small{Proposed (Fading)}  & \small{Zhang \cite{zhang2016capacity}}  &  \small{Sharma} \cite{sharma_globecom}& 	\small{Yu \cite{star}} &  \small{Ma \cite{NOMA_book}} \\ \hline 	\hline
			 $d_{E,\min}$ & 0.8966 &  0.8625 & 1.0171 &   0.9976 &  0.5351 &  0.3883 \\ \hline
		 	$\tau_E$  & 4 & 4 & 2 & 2 & 2 & 4 \\  \hline
			$d_{P,\min}$ & 0.1103 & 0.0595 &  0.0810 &   0.0544 & 0.0379 &   0.0448 \\ \hline
			 $\tau_P$  & 4 & 4 & 4 & 4 & 2 & 2   \\  \hline  \hline
			
	\end{tabular}}
\end{table}
 These KPIs only partially characterize the SCMA system.
The SCMA system is a complicated mult-user scenario where the detection is carried out by the sophisticated MPA.
These KPIs fail to  take the MPA-based detection process  into account.  Thus  they are inadequate to facilitate  a conclusive comparative analysis of various codebooks.
 \begin{table*}[!htbp]
 	\caption{Structure  $\mathbb{S}$ of the codebooks ${\bf{C}}=\left\{{\cal{C}}_1, \ldots, {\cal{C}}_{12}\right\}$  for the $12\times 6$  SCMA system with overloading factor of 200$\%$}
 	\label{table:codebook_structure1}
 	\scalebox{0.75}{$\begin{array}{l}
 		{C_1} = \left\{ {\left[ {\begin{array}{*{20}{c}}
 				{{a_1}}\\
 				{{a_3}}\\
 				0\\
 				\begin{array}{l}
 				0\\
 				0\\
 				0
 				\end{array}
 				\end{array}} \right]\left[ {\begin{array}{*{20}{c}}
 				{{a_2}}\\
 				{{a_4}}\\
 				0\\
 				\begin{array}{l}
 				0\\
 				0\\
 				0
 				\end{array}
 				\end{array}} \right]\left[ {\begin{array}{*{20}{c}}
 				{ - {a_2}}\\
 				{ - {a_4}}\\
 				0\\
 				\begin{array}{l}
 				0\\
 				0\\
 				0
 				\end{array}
 				\end{array}} \right]\left[ {\begin{array}{*{20}{c}}
 				{ - {a_1}}\\
 				{ - {a_3}}\\
 				0\\
 				\begin{array}{l}
 				0\\
 				0\\
 				0
 				\end{array}
 				\end{array}} \right]} \right\}\;{C_2} = \left\{ {\left[ {\begin{array}{*{20}{c}}
 				{{a_3}}\\
 				0\\
 				{{a_5}}\\
 				\begin{array}{l}
 				0\\
 				0\\
 				0
 				\end{array}
 				\end{array}} \right]\left[ {\begin{array}{*{20}{c}}
 				{{a_4}}\\
 				0\\
 				{{a_6}}\\
 				\begin{array}{l}
 				0\\
 				0\\
 				0
 				\end{array}
 				\end{array}} \right]\left[ {\begin{array}{*{20}{c}}
 				{ - {a_4}}\\
 				0\\
 				{ - {a_6}}\\
 				\begin{array}{l}
 				0\\
 				0\\
 				0
 				\end{array}
 				\end{array}} \right]\left[ {\begin{array}{*{20}{c}}
 				{ - {a_3}}\\
 				0\\
 				{ - {a_5}}\\
 				\begin{array}{l}
 				0\\
 				0\\
 				0
 				\end{array}
 				\end{array}} \right]} \right\}\;{C_3} = \left\{ {\left[ {\begin{array}{*{20}{c}}
 				{{a_5}}\\
 				0\\
 				0\\
 				\begin{array}{l}
 				- {a_8}\\
 				\;\;0\\
 				\;\;0
 				\end{array}
 				\end{array}} \right]\left[ {\begin{array}{*{20}{c}}
 				{{a_6}}\\
 				0\\
 				0\\
 				\begin{array}{l}
 				{a_7}\\
 				0\\
 				0
 				\end{array}
 				\end{array}} \right]\left[ {\begin{array}{*{20}{c}}
 				{ - {a_6}}\\
 				0\\
 				0\\
 				\begin{array}{l}
 				- {a_7}\\
 				\;\;0\\
 				\;\;0
 				\end{array}
 				\end{array}} \right]\left[ {\begin{array}{*{20}{c}}
 				{ - {a_5}}\\
 				0\\
 				0\\
 				\begin{array}{l}
 				{a_8}\\
 				0\\
 				0
 				\end{array}
 				\end{array}} \right]} \right\}\\
 		\\
 		\;{C_4} = \left\{ {\left[ {\begin{array}{*{20}{c}}
 				{{a_7}}\\
 				0\\
 				0\\
 				\begin{array}{l}
 				0\\
 				{a_1}\\
 				0
 				\end{array}
 				\end{array}} \right]\left[ {\begin{array}{*{20}{c}}
 				{{a_8}}\\
 				0\\
 				0\\
 				\begin{array}{l}
 				0\\
 				{a_2}\\
 				0
 				\end{array}
 				\end{array}} \right]\left[ {\begin{array}{*{20}{c}}
 				{ - {a_8}}\\
 				0\\
 				0\\
 				\begin{array}{l}
 				0\\
 				- {a_2}\\
 				0
 				\end{array}
 				\end{array}} \right]\left[ {\begin{array}{*{20}{c}}
 				{ - {a_7}}\\
 				0\\
 				0\\
 				\begin{array}{l}
 				\;\;0\\
 				- {a_1}\\
 				\;\;0
 				\end{array}
 				\end{array}} \right]} \right\}{C_5} = \left\{ {\left[ {\begin{array}{*{20}{c}}
 				0\\
 				{{a_5}}\\
 				0\\
 				\begin{array}{l}
 				{a_3}\\
 				\;\;0\\
 				\;\;0
 				\end{array}
 				\end{array}} \right]\left[ {\begin{array}{*{20}{c}}
 				0\\
 				{{a_6}}\\
 				0\\
 				\begin{array}{l}
 				{a_4}\\
 				0\\
 				0
 				\end{array}
 				\end{array}} \right]\left[ {\begin{array}{*{20}{c}}
 				0\\
 				{ - {a_6}}\\
 				0\\
 				\begin{array}{l}
 				- {a_4}\\
 				\;\;0\\
 				\;\;0
 				\end{array}
 				\end{array}} \right]\left[ {\begin{array}{*{20}{c}}
 				0\\
 				{ - {a_5}}\\
 				0\\
 				\begin{array}{l}
 				- {a_3}\\
 				0\\
 				0
 				\end{array}
 				\end{array}} \right]} \right\}\;{C_6} = \left\{ {\left[ {\begin{array}{*{20}{c}}
 				0\\
 				{{a_7}}\\
 				0\\
 				\begin{array}{l}
 				\;\;0\\
 				- {a_4}\\
 				0
 				\end{array}
 				\end{array}} \right]\left[ {\begin{array}{*{20}{c}}
 				0\\
 				{{a_8}}\\
 				0\\
 				\begin{array}{l}
 				0\\
 				{a_3}\\
 				0
 				\end{array}
 				\end{array}} \right]\left[ {\begin{array}{*{20}{c}}
 				0\\
 				{ - {a_8}}\\
 				0\\
 				\begin{array}{l}
 				0\\
 				- {a_3}\\
 				0
 				\end{array}
 				\end{array}} \right]\left[ {\begin{array}{*{20}{c}}
 				0\\
 				{ - {a_7}}\\
 				0\\
 				\begin{array}{l}
 				\;\;0\\
 				{a_4}\\
 				\;\;0
 				\end{array}
 				\end{array}} \right]} \right\}\\
 		\\
 		{C_7} = \left\{ {\left[ {\begin{array}{*{20}{c}}
 				0\\
 				{{a_1}}\\
 				0\\
 				\begin{array}{l}
 				0\\
 				0\\
 				{a_5}
 				\end{array}
 				\end{array}} \right]\left[ {\begin{array}{*{20}{c}}
 				0\\
 				{{a_2}}\\
 				0\\
 				\begin{array}{l}
 				0\\
 				0\\
 				{a_6}
 				\end{array}
 				\end{array}} \right]\left[ {\begin{array}{*{20}{c}}
 				0\\
 				{ - {a_2}}\\
 				0\\
 				\begin{array}{l}
 				\;\;0\\
 				\;\;0\\
 				- {a_6}
 				\end{array}
 				\end{array}} \right]\left[ {\begin{array}{*{20}{c}}
 				0\\
 				{ - {a_1}}\\
 				0\\
 				\begin{array}{l}
 				\;0\\
 				\;0\\
 				- {a_5}
 				\end{array}
 				\end{array}} \right]} \right\}\;{C_8} = \left\{ {\left[ {\begin{array}{*{20}{c}}
 				0\\
 				0\\
 				{{a_7}}\\
 				\begin{array}{l}
 				\;\;0\\
 				0\\
 				{a_1}
 				\end{array}
 				\end{array}} \right]\left[ {\begin{array}{*{20}{c}}
 				0\\
 				0\\
 				{{a_8}}\\
 				\begin{array}{l}
 				0\\
 				0\\
 				{a_2}
 				\end{array}
 				\end{array}} \right]\left[ {\begin{array}{*{20}{c}}
 				0\\
 				0\\
 				{ - {a_8}}\\
 				\begin{array}{l}
 				0\\
 				0\\
 				- {a_2}
 				\end{array}
 				\end{array}} \right]\left[ {\begin{array}{*{20}{c}}
 				0\\
 				0\\
 				{ - {a_7}}\\
 				\begin{array}{l}
 				\;\;0\\
 				\;0\\
 				\; - {a_1}
 				\end{array}
 				\end{array}} \right]} \right\}{C_9} = \left\{ {\left[ {\begin{array}{*{20}{c}}
 				0\\
 				0\\
 				{{a_1}}\\
 				\begin{array}{l}
 				0\\
 				{a_5}\\
 				0
 				\end{array}
 				\end{array}} \right]\left[ {\begin{array}{*{20}{c}}
 				0\\
 				0\\
 				{{a_2}}\\
 				\begin{array}{l}
 				0\\
 				{a_6}\\
 				0
 				\end{array}
 				\end{array}} \right]\left[ {\begin{array}{*{20}{c}}
 				0\\
 				0\\
 				{ - {a_2}}\\
 				\begin{array}{l}
 				\;\;0\\
 				- {a_6}\\
 				0
 				\end{array}
 				\end{array}} \right]\left[ {\begin{array}{*{20}{c}}
 				0\\
 				0\\
 				{ - {a_1}}\\
 				\begin{array}{l}
 				\;0\\
 				- {a_5}\\
 				\;0
 				\end{array}
 				\end{array}} \right]} \right\}\;\;\\
 		\\
 		{C_{10}} = \left\{ {\left[ {\begin{array}{*{20}{c}}
 				0\\
 				0\\
 				{{a_3}}\\
 				\begin{array}{l}
 				0\\
 				0\\
 				{a_1}
 				\end{array}
 				\end{array}} \right]\left[ {\begin{array}{*{20}{c}}
 				0\\
 				0\\
 				{{a_4}}\\
 				\begin{array}{l}
 				0\\
 				0\\
 				{a_2}
 				\end{array}
 				\end{array}} \right]\left[ {\begin{array}{*{20}{c}}
 				0\\
 				0\\
 				{ - {a_4}}\\
 				\begin{array}{l}
 				\;\;0\\
 				\;\;0\\
 				- {a_2}
 				\end{array}
 				\end{array}} \right]\left[ {\begin{array}{*{20}{c}}
 				0\\
 				0\\
 				{ - {a_3}}\\
 				\begin{array}{l}
 				\;0\\
 				\;0\\
 				- {a_1}
 				\end{array}
 				\end{array}} \right]} \right\}{C_{11}} = \left\{ {\left[ {\begin{array}{*{20}{c}}
 				0\\
 				0\\
 				0\\
 				\begin{array}{l}
 				{a_5}\\
 				0\\
 				{a_7}
 				\end{array}
 				\end{array}} \right]\left[ {\begin{array}{*{20}{c}}
 				0\\
 				0\\
 				0\\
 				\begin{array}{l}
 				{a_6}\\
 				0\\
 				{a_8}
 				\end{array}
 				\end{array}} \right]\left[ {\begin{array}{*{20}{c}}
 				0\\
 				0\\
 				0\\
 				\begin{array}{l}
 				- {a_6}\\
 				\;0\\
 				\; - {a_8}
 				\end{array}
 				\end{array}} \right]\left[ {\begin{array}{*{20}{c}}
 				0\\
 				0\\
 				0\\
 				\begin{array}{l}
 				- {a_5}\\
 				- {a_5}\\
 				- {a_7}
 				\end{array}
 				\end{array}} \right]} \right\}{C_{12}} = \left\{ {\left[ {\begin{array}{*{20}{c}}
 				0\\
 				0\\
 				0\\
 				\begin{array}{l}
 				0\\
 				{a_7}\\
 				{a_3}
 				\end{array}
 				\end{array}} \right]\left[ {\begin{array}{*{20}{c}}
 				0\\
 				0\\
 				0\\
 				\begin{array}{l}
 				0\\
 				{a_8}\\
 				{a_4}
 				\end{array}
 				\end{array}} \right]\left[ {\begin{array}{*{20}{c}}
 				0\\
 				0\\
 				0\\
 				\begin{array}{l}
 				\;\;0\\
 				- {a_8}\\
 				- {a_4}
 				\end{array}
 				\end{array}} \right]\left[ {\begin{array}{*{20}{c}}
 				0\\
 				0\\
 				0\\
 				\begin{array}{l}
 				\;0\\
 				- {a_7}\\
 				- {a_3}
 				\end{array}
 				\end{array}} \right]} \right\}
 		\end{array}$}
 \end{table*}
The above codebooks  can be enlarged to build  $J=8$, $K=4$ SCMA systems with 200$\%$ overloading factor. We consider the following factor matrix:
\begin{equation}
\label{fe}
	{\bf{F}}=
	\begin{bmatrix}
	1 & 0 & 1 & 0 & 1 & 0 & 1& 0\\
	0 & 1 & 1 & 0 & 0 & 1 & 1& 0\\
	1 & 0 & 0 & 1 & 0 & 1 & 0& 1\\
	0 & 1 & 0 & 1 & 1 & 0 & 0& 1\\
	\end{bmatrix}.
\end{equation}

\begin{figure}[htb!]
	\centering
	\scalebox{0.8}{\includegraphics[]{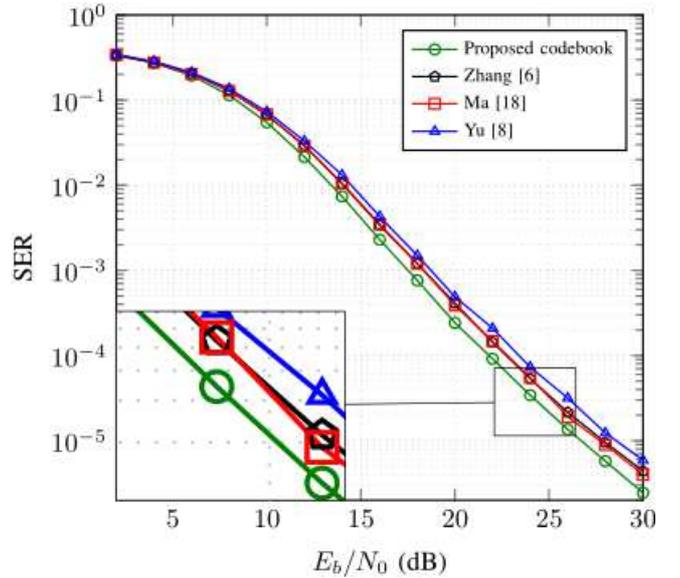}}
	\caption {SER performance of the SCMA system using various codebooks for  $J=8$ and $K=4$  in Rayleigh fading channel.}	 	
	\label{8by4_fading}
\end{figure}

The  $3^{\text{rd}}$ and the $4^{\text{th}}$  columns are repeated as the $8^{\text{th}}$ and $7^{\text{th}}$ ones respectively. 
The constellation points for  ${\cal{C}}_3$ and ${\cal{C}}_4$ are exchanged between the $7^{\text{th}}$  and the $8^{\text{th}}$  users. The SER performances of these codebooks are shown in Fig.~\ref{8by4_fading} for Rayleigh fading channel.  In this case also, the proposed DE-based codebooks yield the best result.

\newpage
\subsection*{\bf{Dedicated SCMA systems with overloading factor of 200$\%$}}

In this section we discuss the construction dedicated codebooks with 200$\%$ overloading factor. The overloading factor for power-domain NOMA is 200$\%$. Therefore, design of 200$\%$ overloaded SCMA codebooks is extremely important in showcasing the relevance of SCMA technique.   Observe  that the factor graph corresponding to the factor matrix in (\ref{fe})  contains  4-cycles.  4-cycles are detrimental for MPA and must be avoided.   Therefore, we consider  an SCMA system of $J=12$ users and $K=6$ resources  with the following factor matrix:  
	\begin{equation}
	\label{expanded_F_r3}
	{\bf{F}}=
\left[ {\begin{array}{*{20}{c}}
	1&1&1&1&0&0&0&0&0&0&0&0\\
	1&0&0&0&1&1&1&0&0&0&0&0\\
	0&1&0&0&0&0&0&1&1&1&0&0\\
	0&0&1&0&1&0&0&1&0&0&1&0\\
	0&0&0&1&0&1&0&0&1&0&0&1\\
	0&0&0&0&0&0&1&0&0&1&1&1
	\end{array}} \right]
	\end{equation}
		\renewcommand{\arraystretch}{1}
		\begin{table*}[!htbp]
			\caption{Codebooks of $12\times6$ SCMA system  optimized for AWGN channel}
			\label{table:codebooks_AWGN}
			\centering
			\scalebox{0.65}{
				$
				\begin{gathered}
				{\cal{C}}_1=\left\{\begin{bmatrix}
				0.0140 - 0.0157i \\
				-0.3296 + 0.9439i \\
				0 \\
				0\\
				0\\
				0
				\end{bmatrix}
				\begin{bmatrix}
				 0.2553 + 0.4557i \\
			-0.8508 + 0.0578i \\
				0 \\
				0 \\
				0\\
				0
				\end{bmatrix}
				\begin{bmatrix}
				-0.2553 - 0.4557i \\
				 0.8508 - 0.0578i  \\
				0 \\
				0 \\
				0\\
				0
				\end{bmatrix}
				\begin{bmatrix}
				-0.0140 + 0.0157i \\
				 0.3296 - 0.9439i\\
				0 \\
				0\\
				0\\
				0
				\end{bmatrix}  \right\} 
				{\cal{C}}_2=\left\{\begin{bmatrix}
				-0.3151 + 0.9024i \\
				0\\
				0.2879 + 0.0592i\\
				0\\
				0\\
				0
				\end{bmatrix}
				\begin{bmatrix}
				-0.8064 + 0.0547i \\
				0\\
				-0.3961 + 0.4357i\\
				0\\
				0\\
				0
				\end{bmatrix}
				\begin{bmatrix}
				0.8064 - 0.0547i  \\
				0\\
				0.3961 - 0.4357i \\
				0\\
				0\\
				0
				\end{bmatrix}
				\begin{bmatrix}
			0.3151 - 0.9024i\\
				0\\
			-0.2879 - 0.0592i\\
				0\\
				0\\
				0
				\end{bmatrix}   \right\} \\
				{\cal{C}}_3=\left\{\begin{bmatrix}
				0.3350 + 0.0688i \\
				0\\
				0\\
				-0.2762 - 0.8982i\\
				0\\
				0
				\end{bmatrix}
				\begin{bmatrix}
				-0.3550 + 0.3905i \\
				0\\
				0\\
				 0.7463 + 0.4056i\\
				0\\
				0
				\end{bmatrix}
				\begin{bmatrix}
				 0.3550 - 0.3905i \\
				0\\
				0\\
			-0.7463 - 0.4056i \\
				0\\
				0
				\end{bmatrix}
				\begin{bmatrix}
				-0.3350 - 0.0688i \\
				0\\
				0\\
				0.2762 + 0.8982i\\
				0\\
				0
				\end{bmatrix}   \right\}  \quad 
				{\cal{C}}_4=\left\{\begin{bmatrix}
			0.8784 + 0.4774i\\
				0\\
				0\\
				0\\
			0.0140 - 0.0157i\\
				0
				\end{bmatrix}
				\begin{bmatrix}
			0.2500 + 0.8129i\\
				0\\
				0\\
				0\\
				0.2571 + 0.4589i\\
				0
				\end{bmatrix}
				\begin{bmatrix}
				-0.2500 - 0.8129i\\
				0\\
				0\\
				0\\
				-0.2571 - 0.4589i \\
				0
				\end{bmatrix}
				\begin{bmatrix}
				-0.8784 - 0.4774i\\
				0\\
				0\\
				0\\
			-0.0140 + 0.0157i\\
				0
				\end{bmatrix}   \right\} \\
				{\cal{C}}_5=\left\{\begin{bmatrix}
				0\\
				0.2879 + 0.0592i \\
				0\\
			-0.3151 + 0.9024i\\
				0\\
				0
				\end{bmatrix}
				\begin{bmatrix}
				0\\
			-0.3961 + 0.4357i \\
				0\\
				-0.8064 + 0.0547i\\
				0\\
				0
				\end{bmatrix}
				\begin{bmatrix}
				0\\
			0.3961 - 0.4357i\\
				0\\
			0.8064 - 0.0547i\\
				0\\
				0
				\end{bmatrix}
				\begin{bmatrix}
				0\\
			-0.2879 - 0.0592i \\
				0\\
			 0.3151 - 0.9024i\\
				0\\
				0
				\end{bmatrix}   \right\} \quad
				{\cal{C}}_6=\left\{\begin{bmatrix}
				0\\
				0.6685 + 0.3633i\\
				0\\
				0\\
				0.6474 - 0.0439i \\
				0
				\end{bmatrix}
				\begin{bmatrix}
				0\\
			 0.1897 + 0.6168i \\
				0\\
				0\\
				 -0.2519 + 0.7212i  \\
				0
				\end{bmatrix}
				\begin{bmatrix}
				0\\
				-0.1897 - 0.6168i\\
				0\\
				0\\
				0.2519 - 0.7212i  \\
				0
				\end{bmatrix}
				\begin{bmatrix}
				0\\
				-0.6685 - 0.3633i\\
				0\\
				0\\
				-0.6474 + 0.0439i \\
				0
				\end{bmatrix}   \right\}\\
				{\cal{C}}_7=\left\{\begin{bmatrix}
				0\\
			0.0456 - 0.0509i\\
				0\\
				0\\
				0 \\
				0.9772 + 0.2008i 
				\end{bmatrix}
				\begin{bmatrix}
				0\\
				0.3145 + 0.5615i \\
				0\\
				0\\
				0\\
				-0.5148 + 0.5664i   
				
				\end{bmatrix}
				\begin{bmatrix}
				0\\
				-0.3145 - 0.5615i \\
				0\\
				0\\
				0\\
				0.5148 - 0.5664i  
				
				\end{bmatrix}
				\begin{bmatrix}
				0\\
			-0.0456 + 0.0509i\\
				0\\
				0\\
				0 \\
			-0.9772 - 0.2008i
				\end{bmatrix}   \right\}	\quad
				{\cal{C}}_8=\left\{\begin{bmatrix}
				0\\
				0\\
				0.8784 + 0.4774i\\
				0.0140 - 0.0157i\\
				0 \\
				0 
				\end{bmatrix}
				\begin{bmatrix}
				0\\
				0\\
				0.2500 + 0.8129i\\
				0.2571 + 0.4589i \\
				0 \\
				0 
				\end{bmatrix}
				\begin{bmatrix}
				0\\
				0\\
				-0.2500 - 0.8129i\\
			-0.2571 - 0.4589i \\
				0 \\
				0 
				\end{bmatrix}
				\begin{bmatrix}
				0\\
				0\\
				-0.8784 - 0.4774i\\
			-0.0140 + 0.0157i\\
				0 \\
				0 
				\end{bmatrix}   \right\}	\\
				{\cal{C}}_9=\left\{\begin{bmatrix}
				0\\
				0\\
				0.0456 - 0.0509i\\
				0\\
				0.9772 + 0.2008i \\
				0 
				\end{bmatrix}
				\begin{bmatrix}
				0\\
				0\\
				0.3145 + 0.5615i\\
				0\\
				-0.5148 + 0.5664i \\
				0 
				\end{bmatrix}
				\begin{bmatrix}
				0\\
				0\\
				-0.3145 - 0.5615i\\
				0\\
				0.5148 - 0.5664i \\
				0   
				\end{bmatrix}
				\begin{bmatrix}
				0\\
				0\\
				-0.0456 + 0.0509i\\
				0\\
				-0.9772 - 0.2008i \\
				0 
				\end{bmatrix}   \right\}	\quad
				{\cal{C}}_{10}=\left\{\begin{bmatrix}
				0\\
				0\\
				-0.3296 + 0.9439i \\
				0\\
				0 \\
				0.0140 - 0.0157i   
				\end{bmatrix}
				\begin{bmatrix}
				0\\
				0\\
				-0.8508 + 0.0578i\\
				0\\
				0 \\
			 0.2553 + 0.4557i 
				\end{bmatrix}
				\begin{bmatrix}
				0\\
				0\\
			0.8508 - 0.0578i\\
				0\\
				0 \\
			-0.2553 - 0.4557i
				\end{bmatrix}
				\begin{bmatrix}
				0\\
				0\\
			0.3296 - 0.9439i \\
				0\\
				0 \\
			-0.0140 + 0.0157i
				\end{bmatrix}   \right\}\\
				{\cal{C}}_{11}=\left\{\begin{bmatrix}
				0\\
				0\\
				0 \\
			 0.2879 + 0.0592i\\
				0 \\
			 0.8398 + 0.4564i 
				\end{bmatrix}
				\begin{bmatrix}
				0\\
				0\\
				0 \\
			-0.3985 + 0.4384i \\
				0 \\
			0.2368 + 0.7700i 
				\end{bmatrix}
				\begin{bmatrix}
				0\\
				0\\
				0 \\
			0.3985 - 0.4384i \\
				0 \\
			-0.2368 - 0.7700i 
				\end{bmatrix}		
				\begin{bmatrix}
				0\\
				0\\
				0 \\
			 -0.2879 - 0.0592i\\
				0 \\
			-0.8398 - 0.4564i
				\end{bmatrix}   \right\} \quad 
				{\cal{C}}_{12}=\left\{\begin{bmatrix}
				0\\
				0\\
				0 \\
				0\\
				0.6213 + 0.3376i  \\
			-0.2331 + 0.6676i  
				\end{bmatrix}
				\begin{bmatrix}
				0\\
				0\\
				0 \\
				0\\
				0.2068 + 0.6727i   \\
			-0.7088 + 0.0481i  
				\end{bmatrix}
				\begin{bmatrix}
				0\\
				0\\
				0 \\
				0\\
				-0.2068 - 0.6727i   \\
				0.7088 - 0.0481i
				\end{bmatrix}		
				\begin{bmatrix}
				0\\
				0\\
				0 \\
				0\\
				-0.6213 - 0.3376i \\
			0.2331 - 0.6676i 
				\end{bmatrix}   \right\}
				\end{gathered}
				$}
			
		\end{table*}
				\begin{table*}[!htbp]
					\caption{Codebooks of $12\times6$ SCMA system  optimized for Rayleigh fading  channel}
					\label{table:codebooks_ray}
					\centering
					\scalebox{0.65}{
						$
						\begin{gathered}
						{\cal{C}}_1=\left\{\begin{bmatrix}
						0.7435 + 0.0752i \\
					 -0.6626 - 0.0500i\\
						0 \\
						0\\
						0\\
						0
						\end{bmatrix}
						\begin{bmatrix}
					-0.0059 + 0.7487i \\
					-0.0691 - 0.6592i   \\
						0 \\
						0 \\
						0\\
						0
						\end{bmatrix}
						\begin{bmatrix}
						0.0059 - 0.7487i \\
					0.0691 + 0.6592i  \\
						0 \\
						0 \\
						0\\
						0
						\end{bmatrix}
						\begin{bmatrix}
						-0.7435 - 0.0752i \\
						 0.6626 + 0.0500i \\
						0 \\
						0\\
						0\\
						0
						\end{bmatrix}  \right\} 
						{\cal{C}}_2=\left\{\begin{bmatrix}
						-0.6592 - 0.0498i \\
						0\\
						0.3767 + 0.6489i\\
						0\\
						0\\
						0
						\end{bmatrix}
						\begin{bmatrix}
						-0.0673 - 0.6420i  \\
						0\\
						-0.6278 + 0.4350i\\
						0\\
						0\\
						0
						\end{bmatrix}
						\begin{bmatrix}
					0.0673 + 0.6420i   \\
						0\\
					0.6278 - 0.4350i\\
						0\\
						0\\
						0
						\end{bmatrix}
						\begin{bmatrix}
					0.6592 + 0.0498i\\
						0\\
						-0.3767 - 0.6489i\\
						0\\
						0\\
						0
						\end{bmatrix}   \right\} \\
						{\cal{C}}_3=\left\{\begin{bmatrix}
						0.3820 + 0.6581i \\
						0\\
						0\\
						0.0430 - 0.6474i\\
						0\\
						0
						\end{bmatrix}
						\begin{bmatrix}
						-0.6315 + 0.4376i \\
						0\\
						0\\
						0.6382 + 0.0491i\\
						0\\
						0
						\end{bmatrix}
						\begin{bmatrix}
						 0.6315 - 0.4376i \\
						0\\
						0\\
						-0.6382 - 0.0491i\\
						0\\
						0
						\end{bmatrix}
						\begin{bmatrix}
						-0.3820 - 0.6581i\\
						0\\
						0\\
						-0.0430 + 0.6474i\\
						0\\
						0
						\end{bmatrix}   \right\}  \quad 
						{\cal{C}}_4=\left\{\begin{bmatrix}
						0.6563 + 0.0505i\\
						0\\
						0\\
						0\\
						0.7490 + 0.0758i\\
						0
						\end{bmatrix}
						\begin{bmatrix}
						-0.0432 + 0.6500i\\
						0\\
						0\\
						0\\
						-0.0060 + 0.7586i\\
						0
						\end{bmatrix}
						\begin{bmatrix}
					0.0432 - 0.6500i\\
						0\\
						0\\
						0\\
						0.0060 - 0.7586i\\
						0
						\end{bmatrix}
						\begin{bmatrix}
						-0.6563 - 0.0505i\\
						0\\
						0\\
						0\\
						-0.7490 - 0.0758i\\
						0
						\end{bmatrix}   \right\} \\
						{\cal{C}}_5=\left\{\begin{bmatrix}
						0\\
						0.3767 + 0.6489i \\
						0\\
						-0.6592 - 0.0498i\\
						0\\
						0
						\end{bmatrix}
						\begin{bmatrix}
						0\\
						-0.6278 + 0.4350i \\
						0\\
						-0.0673 - 0.6420i \\
						0\\
						0
						\end{bmatrix}
						\begin{bmatrix}
						0\\
						0.6278 - 0.4350i \\
						0\\
						0.0673 + 0.6420i \\
						0\\
						0
						\end{bmatrix}
						\begin{bmatrix}
						0\\
					-0.3767 - 0.6489i \\
						0\\
						0.6592 + 0.0498i\\
						0\\
						0
						\end{bmatrix}   \right\} \quad
						{\cal{C}}_6=\left\{\begin{bmatrix}
						0\\
					0.7000 + 0.0538i \\
						0\\
						0\\
					 0.0743 + 0.7082i \\
						0
						\end{bmatrix}
						\begin{bmatrix}
						0\\
						-0.0461 + 0.6938i\\
						0\\
						0\\
						-0.7166 - 0.0541i  \\
						0
						\end{bmatrix}
						\begin{bmatrix}
						0\\
						 0.0461 - 0.6938i\\
						0\\
						0\\
					0.7166 + 0.0541i  \\
						0
						\end{bmatrix}
						\begin{bmatrix}
						0\\
						-0.7000 - 0.0538i\\
						0\\
						0\\
					-0.0743 - 0.7082i \\
						0
						\end{bmatrix}   \right\}\\
						{\cal{C}}_7=\left\{\begin{bmatrix}
						0\\
					0.7003 + 0.0709i\\
						0\\
						0\\
						0 \\
						0.3566 + 0.6143i
						\end{bmatrix}
						\begin{bmatrix}
						0\\
						-0.0054 + 0.6905i \\
						0\\
						0\\
						0\\
						-0.5945 + 0.4120i   
						
						\end{bmatrix}
						\begin{bmatrix}
						0\\
						0.0054 - 0.6905i\\
						0\\
						0\\
						0\\
						0.5945 - 0.4120i
						
						\end{bmatrix}
						\begin{bmatrix}
						0\\
						-0.7003 - 0.0709i\\
						0\\
						0\\
						0 \\
					-0.3566 - 0.6143i
						\end{bmatrix}   \right\}	\quad
						{\cal{C}}_8=\left\{\begin{bmatrix}
						0\\
						0\\
					  0.6563 + 0.0505i \\
					0.7490 + 0.0758i\\
						0 \\
						0 
						\end{bmatrix}
						\begin{bmatrix}
						0\\
						0\\
					-0.0432 + 0.6500i\\
					-0.0060 + 0.7586i \\
						0 \\
						0 
						\end{bmatrix}
						\begin{bmatrix}
						0\\
						0\\
						0.0432 - 0.6500i\\
						0.0060 - 0.7586i \\
						0 \\
						0 
						\end{bmatrix}
						\begin{bmatrix}
						0\\
						0\\
						 -0.6563 - 0.0505i\\
					-0.7490 - 0.0758i\\
						0 \\
						0 
						\end{bmatrix}   \right\}	\\
						{\cal{C}}_9=\left\{\begin{bmatrix}
						0\\
						0\\
						0.7003 + 0.0709i\\
						0\\
						0.3566 + 0.6143i \\
						0 
						\end{bmatrix}
						\begin{bmatrix}
						0\\
						0\\
						 -0.0054 + 0.6905i\\
						0\\
					-0.5945 + 0.4120i \\
						0 
						\end{bmatrix}
						\begin{bmatrix}
						0\\
						0\\
						0.0054 - 0.6905i\\
						0\\
						0.5945 - 0.4120i \\
						0   
						\end{bmatrix}
						\begin{bmatrix}
						0\\
						0\\
						-0.7003 - 0.0709i\\
						0\\
					-0.3566 - 0.6143i \\
						0 
						\end{bmatrix}   \right\}	\quad
						{\cal{C}}_{10}=\left\{\begin{bmatrix}
						0\\
						0\\
						-0.6626 - 0.0500i \\
						0\\
						0 \\
					0.7435 + 0.0752i
						\end{bmatrix}
						\begin{bmatrix}
						0\\
						0\\
					-0.0691 - 0.6592i  \\
						0\\
						0 \\
					-0.0059 + 0.7487i 
						\end{bmatrix}
						\begin{bmatrix}
						0\\
						0\\
					0.0691 + 0.6592i \\
						0\\
						0 \\
					0.0059 - 0.7487i 
						\end{bmatrix}
						\begin{bmatrix}
						0\\
						0\\
						0.6626 + 0.0500i \\
						0\\
						0 \\
					-0.7435 - 0.0752i
						\end{bmatrix}   \right\}\\
						{\cal{C}}_{11}=\left\{\begin{bmatrix}
						0\\
						0\\
						0 \\
						 0.3794 + 0.6536i\\
						0 \\
						 0.6529 + 0.0502i  
						\end{bmatrix}
						\begin{bmatrix}
						0\\
						0\\
						0 \\
						-0.6356 + 0.4405i \\
						0 \\
						-0.0421 + 0.6326i
						\end{bmatrix}
						\begin{bmatrix}
						0\\
						0\\
						0 \\
						0.6356 - 0.4405i \\
						0 \\
						0.0421 - 0.6326i 
						\end{bmatrix}		
						\begin{bmatrix}
						0\\
						0\\
						0 \\
						-0.3794 - 0.6536i\\
						0 \\
						-0.6529 - 0.0502i 
						\end{bmatrix}   \right\} \quad 
						{\cal{C}}_{12}=\left\{\begin{bmatrix}
						0\\
						0\\
						0 \\
						0\\
						0.6991 + 0.0538i  \\
						-0.7110 - 0.0537i  
						\end{bmatrix}
						\begin{bmatrix}
						0\\
						0\\
						0 \\
						0\\
					-0.0462 + 0.6947i   \\
					-0.0749 - 0.7139i   
						\end{bmatrix}
						\begin{bmatrix}
						0\\
						0\\
						0 \\
						0\\
					 0.0462 - 0.6947i   \\
					0.0749 + 0.7139i
						\end{bmatrix}		
						\begin{bmatrix}
						0\\
						0\\
						0 \\
						0\\
					-0.6991 - 0.0538i  \\
					0.7110 + 0.0537i
						\end{bmatrix}   \right\}
						\end{gathered}
						$}
					
				\end{table*}
Observe from (\ref{expanded_F_r3})  that the girth of the factor graph is 6 as it doesn't contain any cycles of length 4.  
The structure of the codebooks are shown in TABLE~\ref{table:codebook_structure1}. The complex variables are $a_1, a_2, \ldots , a_8$.  The number of real variables is $D=16$. Using Algorithm~\ref{algo:codebook_DE},  the optimum values of $a_i$s are obtained.  The resulting codebooks  for AWGN and Rayleigh fading channel  are  shown in TABLE~\ref{table:codebooks_AWGN}  and TABLE~\ref{table:codebooks_ray} respectively. 
\begin{figure}[htb!]
	\centering
	\scalebox{0.7}{\includegraphics[]{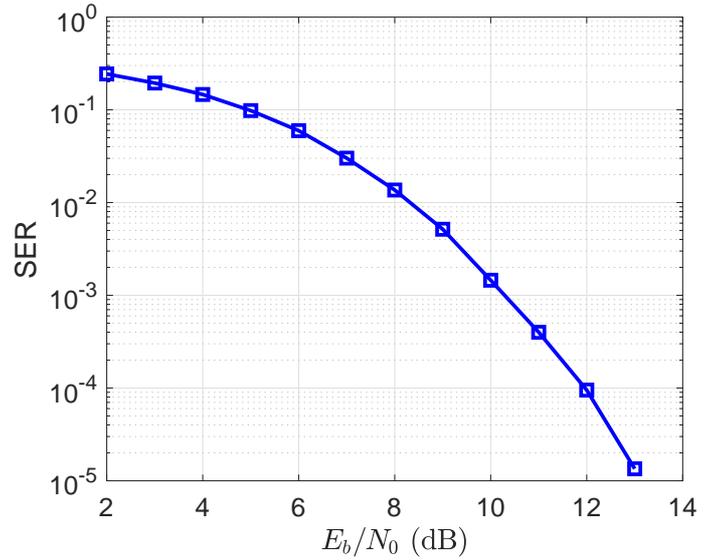}}
	\caption {SER performance of the SCMA system  (TABLE~\ref{table:codebooks_AWGN}) for  $J=12$ and $K=6$    in AWGN channel.}	 	
	\label{12by6_awgn}
\end{figure}
\begin{figure}[htb!]
	\centering
	\scalebox{0.7}{\includegraphics[]{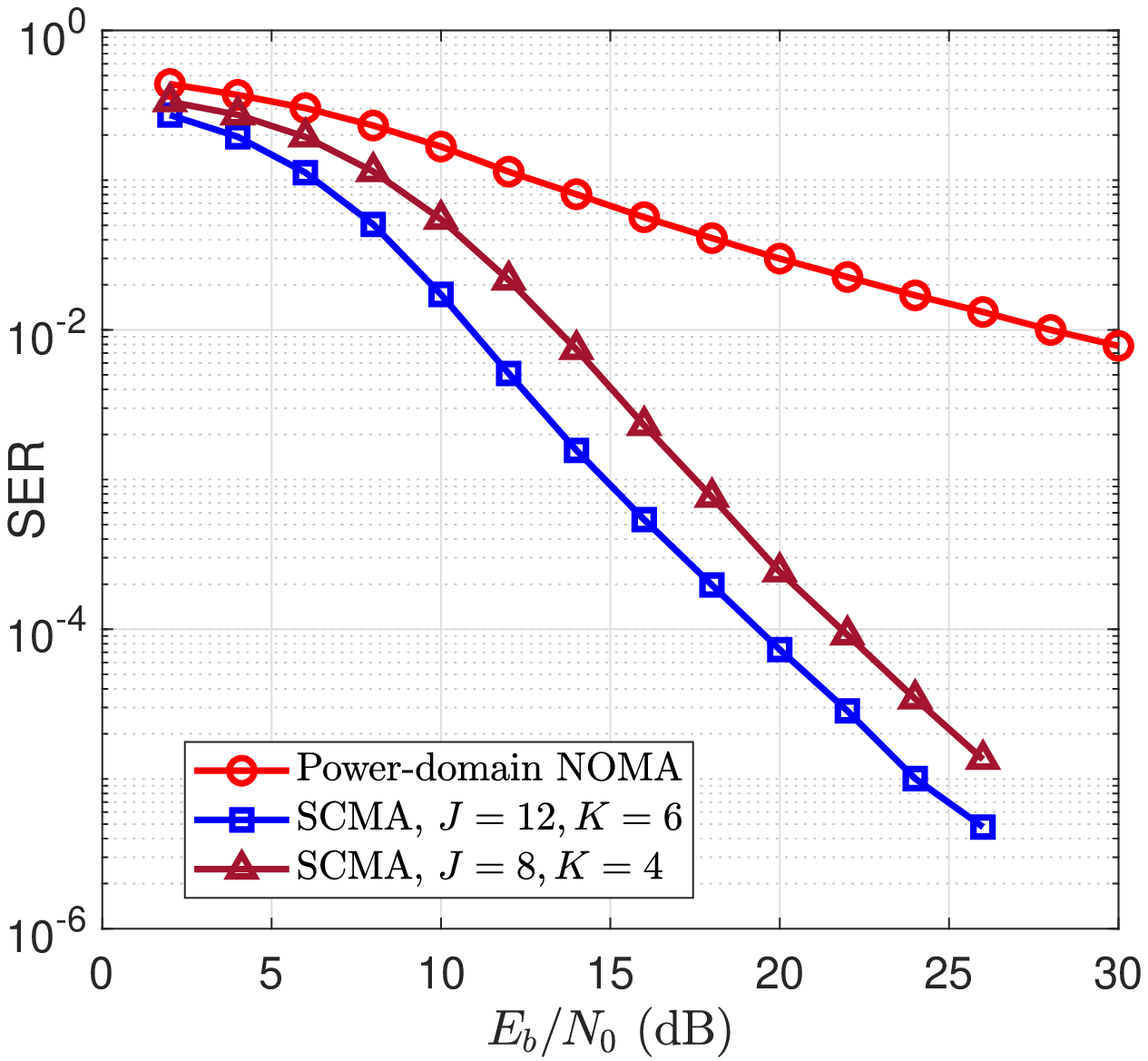}}
	\caption {SER performance of the SCMA system  (TABLE~\ref{table:codebooks_ray}) for  $J=12$ and $K=6$    in Rayleigh fading  channel.}	 	
	\label{12by6_ray}
\end{figure}
Fig.~\ref{12by6_awgn} and  Fig.~\ref{12by6_ray} show the SER of the proposed SCMA codebooks optimized for the  AWGN channel (TABLE~\ref{table:codebooks_AWGN})  and for the Rayleigh fading channel (TABLE~\ref{table:codebooks_ray})  respectively.  Note that the overloading factor of power-domain NOMA is 200$\%$.   The SER performance of the power-domain NOMA  is also shown in Fig.~\ref{12by6_ray}.   The power allocation is carried out by following the maximin
fairness (MMF) \cite{NOMA_book} criterion.   As per the MMF criterion, the optimum  power splits $(P_1,P_2)$ amongst the near and the far users are obtained as $(0.29,0.71), (0.27, 0.73), (0.25, 0.75),$ $ (0.23, 0.77), (0.2, 0.8),  (0.17,83),    (0.15,85),(0.13,87),$    $(0.1, 0.9),     (0.09,0.91), (0.07,0.93), (0.06, 0.94),     (0.05,0.95)$, $    (0.04, 0.96), (0.03,0.97)$ at the SNR points considered in Fig.~\ref{12by6_ray}.  Observe that the  $12\times6$ SCMA system  clearly outperforms the power-domain NOMA system under the same overloading  factor of 200$\%$.  Fig.~\ref{12by6_ray}  also shows the SER performance of the $8\times 4$  system derived from the proposed $6\times 4$  system  shown in TABLE~\ref{table:fading}.  The performance of the $8\times 4$  system is inferior to that of the $12\times 6$  system as the  factor graph  of the former contains 4-cycles.  
\section{Conclusions}  \label{sec::conc}
This paper presented a method to design the codebooks for an SCMA system. The constellation points are designed with the objective of minimizing the SER.  The SER is considered as it  directly reflects the effectiveness of the  system to detect a user's data in interference-limited environment. First the structure of the codebooks is fixed using  a finite number of complex numbers. The minimization of the SER over these variables  is accomplished with the help of DE. The optimum complex numbers are then used to form the desired codebooks. It is found that the codebook-design task is a channel-dependent affairs. The  SER performance of the proposed codebooks for the AWGN and the fading channels are compared with those of other existing codebooks in literature. This comparison established the superiority of the proposed method  over others. 


\bibliographystyle{ieeetran}
\footnotesize
\bibliography{references_SCMA}

\begin{thebibliography}{10}
\providecommand{\url}[1]{#1}
\csname url@samestyle\endcsname
\providecommand{\newblock}{\relax}
\providecommand{\bibinfo}[2]{#2}
\providecommand{\BIBentrySTDinterwordspacing}{\spaceskip=0pt\relax}
\providecommand{\BIBentryALTinterwordstretchfactor}{4}
\providecommand{\BIBentryALTinterwordspacing}{\spaceskip=\fontdimen2\font plus
\BIBentryALTinterwordstretchfactor\fontdimen3\font minus
  \fontdimen4\font\relax}
\providecommand{\BIBforeignlanguage}[2]{{%
\expandafter\ifx\csname l@#1\endcsname\relax
\typeout{** WARNING: IEEEtran.bst: No hyphenation pattern has been}%
\typeout{** loaded for the language `#1'. Using the pattern for}%
\typeout{** the default language instead.}%
\else
\language=\csname l@#1\endcsname
\fi
#2}}
\providecommand{\BIBdecl}{\relax}
\BIBdecl

\bibitem{dai2015non}
L.~Dai, B.~Wang, Y.~Yuan, S.~Han, I.~Chih-Lin, and Z.~Wang, ``Non-orthogonal
  multiple access for 5{G}: solutions, challenges, opportunities, and future
  research trends,'' \emph{IEEE Communications Magazine}, vol.~53, no.~9, pp.
  74--81, 2015.

\bibitem{hoshyar_lds}
R.~Hoshyar, F.~P. Wathan, and R.~Tafazolli, ``Novel low-density signature for
  synchronous {CDMA} systems over {AWGN} channel,'' \emph{IEEE Transactions on
  Signal Processing}, vol.~56, no.~4, pp. 1616--1626, April 2008.

\bibitem{nikopour2013}
H.~Nikopour and H.~Baligh, ``Sparse code multiple access,'' in \emph{2013 IEEE
  24th Annual International Symposium on Personal, Indoor, and Mobile Radio
  Communications (PIMRC)}, Sept 2013, pp. 332--336.

\bibitem{spa}
F.~R. Kschischang, B.~J. Frey, and H.~. Loeliger, ``Factor graphs and the
  sum-product algorithm,'' \emph{IEEE Transactions on Information Theory},
  vol.~47, no.~2, pp. 498--519, Feb 2001.

\bibitem{taherzdeh2014}
M.~{Taherzadeh}, H.~{Nikopour}, A.~{Bayesteh}, and H.~{Baligh}, ``S{CMA}
  codebook design,'' in \emph{2014 IEEE 80th Vehicular Technology Conference
  (VTC2014-Fall)}, Sep. 2014, pp. 1--5.

\bibitem{zhang2016capacity}
S.~Zhang, K.~Xiao, B.~Xiao, Z.~Chen, B.~Xia, D.~Chen, and S.~Ma, ``A
  capacity-based codebook design method for sparse code multiple access
  systems,'' in \emph{2016 8th International Conference on Wireless
  Communications Signal Processing (WCSP)}, Oct 2016, pp. 1--5.

\bibitem{alam2017designing}
M.~Alam and Q.~Zhang, ``Designing optimum mother constellation and codebooks
  for {SCMA},'' in \emph{IEEE International Conference on Communications
  (ICC)}.\hskip 1em plus 0.5em minus 0.4em\relax IEEE, 2017, pp. 1--6.

\bibitem{star}
L.~{Yu}, X.~{Lei}, P.~{Fan}, and D.~{Chen}, ``An optimized design of {SCMA}
  codebook based on star-{QAM} signaling constellations,'' in \emph{2015
  International Conference on Wireless Communications Signal Processing
  (WCSP)}, Oct 2015, pp. 1--5.

\bibitem{Bao_2018TCOM}
J.~{Bao}, Z.~{Ma}, M.~{Xiao}, T.~A. {Tsiftsis}, and Z.~{Zhu}, ``Bit-interleaved
  coded {SCMA} with iterative multiuser detection: Multidimensional
  constellations design,'' \emph{IEEE Transactions on Communications}, vol.~66,
  no.~11, pp. 5292--5304, Nov 2018.

\bibitem{sharma_globecom}
S.~{Sharma}, K.~{Deka}, V.~{Bhatia}, and A.~{Gupta}, ``S{CMA} codebook based on
  optimization of mutual information and shaping gain,'' in \emph{2018 IEEE
  Globecom Workshops (GC Wkshps)}, Dec 2018, pp. 1--6.

\bibitem{kim_deep}
M.~{Kim}, N.~{Kim}, W.~{Lee}, and D.~{Cho}, ``{Deep Learning-Aided SCMA},''
  \emph{IEEE Communications Letters}, vol.~22, no.~4, pp. 720--723, April 2018.

\bibitem{Storn1997}
\BIBentryALTinterwordspacing
R.~Storn and K.~Price, ``{Differential Evolution -- A Simple and Efficient
  Heuristic for global Optimization over Continuous Spaces},'' \emph{Journal of
  Global Optimization}, vol.~11, no.~4, pp. 341--359, Dec 1997. [Online].
  Available: \url{https://doi.org/10.1023/A:1008202821328}
\BIBentrySTDinterwordspacing

\bibitem{DE_2011}
S.~{Das} and P.~N. {Suganthan}, ``Differential evolution: A survey of the
  state-of-the-art,'' \emph{IEEE Transactions on Evolutionary Computation},
  vol.~15, no.~1, pp. 4--31, Feb 2011.

\bibitem{boutros98}
J.~{Boutros} and E.~{Viterbo}, ``Signal space diversity: a power- and
  bandwidth-efficient diversity technique for the rayleigh fading channel,''
  \emph{IEEE Transactions on Information Theory}, vol.~44, no.~4, pp.
  1453--1467, July 1998.

\bibitem{codebook_design_survey}
M.~{Vameghestahbanati}, I.~D. {Marsland}, R.~H. {Gohary}, and
  H.~{Yanikomeroglu}, ``Multidimensional constellations for uplink {SCMA}
  systems, a comparative study,'' \emph{IEEE Communications Surveys Tutorials},
  vol.~21, no.~3, pp. 2169--2194, thirdquarter 2019.

\bibitem{latin}
J.~Denes and A.~D. Keedwell, \emph{{Latin squares: New developments in the
  theory and applications}}.\hskip 1em plus 0.5em minus 0.4em\relax Elsevier,
  1991.

\bibitem{xiao_capacity}
K.~{Xiao}, B.~{Xia}, Z.~{Chen}, B.~{Xiao}, D.~{Chen}, and S.~{Ma}, ``On
  capacity-based codebook design and advanced decoding for sparse code multiple
  access systems,'' \emph{IEEE Transactions on Wireless Communications},
  vol.~17, no.~6, pp. 3834--3849, June 2018.

\bibitem{NOMA_book}
{Mojtaba Vaezi, Zhiguo Ding and H. Vincent Poor }, Ed., \emph{{Multiple Access
  Techniques for 5G Wireless Networks and Beyond}}.\hskip 1em plus 0.5em minus
  0.4em\relax Springer International, 2019.

\bibitem{DE_price_20}
K.~Price, R.~Storm, and J.~Lampinen, \emph{{Differential Evolution}}.\hskip 1em
  plus 0.5em minus 0.4em\relax Springer-Verlag Berlin Heidelberg, 2005.

\bibitem{qing_DE}
{Anyong Qing}, \emph{{Differential Evolution: Fundamentals and Applications in
  Electrical Engineering}}.\hskip 1em plus 0.5em minus 0.4em\relax Wiley 
  IEEE, 2009.

\end{thebibliography}

\end{document}